\documentclass[aps,prl,twocolumn,scriptaddress,superscriptaddress]{revtex4-2}
\usepackage{graphicx,amsmath,amsfonts,amssymb,txfonts,multirow,subfigure,tikz}
\usepackage{xcolor}
\usepackage[colorlinks=false]{hyperref}
\usetikzlibrary{decorations.pathmorphing, decorations.markings, arrows.meta}

\begin{document}
	\newcommand{\fig}[2]{\includegraphics[width=#1]{#2}}
	\newcommand{\la}{{\langle}}
	\newcommand{\ra}{{\rangle}}
	\newcommand{\dg}{{\dagger}}
	\newcommand{\upa}{{\uparrow}}
	\newcommand{\dna}{{\downarrow}}
	\newcommand{\ab}{{\alpha\beta}}
	\newcommand{\ias}{{i\alpha\sigma}}
	\newcommand{\ibs}{{i\beta\sigma}}
	\newcommand{\hH}{\hat{H}}
	\newcommand{\hn}{\hat{n}}
	\newcommand{\hc}{{\hat{\chi}}}
	\newcommand{\hU}{{\hat{U}}}
	\newcommand{\hV}{{\hat{V}}}
	\newcommand{\br}{{\bf r}}
	\newcommand{\bk}{{{\bf k}}}
	\newcommand{\bq}{{{\bf q}}}
	\newcommand{\mr}{\mathrm}
	\def\gsim{~\rlap{$>$}{\lower 1.0ex\hbox{$\sim$}}}
	\setlength{\unitlength}{1mm}
	
	\title{Helical Rashba--exchange gauge field drives a uniaxial pair density wave in EuRbFe$_4$As$_4$}
	
	\author{Pengfei Li}
	\affiliation{Department of Physics, The Hong Kong University of Science and Technology, Clear Water Bay, Kowloon, Hong Kong, China}
	\author{Yi Zhou}
	\affiliation{Institute of Physics, Chinese Academy of Sciences, Beijing 100190, China}
	
	\date{\today}
	
	\begin{abstract}
		The recent discovery of an intrinsic, zero-field pair density wave (PDW) in the iron-pnictide superconductor EuRbFe$_4$As$_4$ poses a fundamental puzzle: how does a unidirectional, nanometer-scale superconducting modulation arise spontaneously below the magnetic ordering temperature? Here we show that the interplay of Rashba spin--orbit coupling---induced by the locally non-centrosymmetric FeAs layers---and the period-four helical Eu$^{2+}$ exchange field generates a layer-rotating effective $U(1)$ gauge field for the Cooper pairs. Because this gauge field shares the symmetry of the Fe $3d_{xz}/3d_{yz}$ orbital doublet, it drives an orbital-selective, finite-momentum pairing instability. Using a Ginzburg--Landau theory on the magnetic unit cell, we demonstrate that this mechanism naturally stabilizes a strictly uniaxial Bloch superconducting state at the experimentally observed wavelength, accompanied by spontaneous interlayer loop currents accessible to muon-spin relaxation or scanning SQUID microscopy.
	\end{abstract}
	
	\maketitle
	
\textbf{Introduction.}  
The pair density wave (PDW) is a superconducting state in which Cooper pairs condense with finite center-of-mass momentum, spontaneously breaking translational symmetry and producing a spatially modulated order parameter even at zero applied field~\cite{PDW-Kivelson2007,Agterberg2008NPdislocations,Berg-cuprate-NJP2009,Berg-PRL2010,Fradkin-PRB2012,cho2012superconductivity,M-F_model_2,PALee-PRX2014,maciejko2014weyl,YXWang-PRL2015,SKJian-PRL2015,SKJian-PRL2017,YXWang-PRB2018,Kim-SciAdv2019,ZYHan-PRL2020,LFu-PRB2020,zhou2021chern,Huang-npjQM2022,Yao-arX2022}. As a parent phase, it can give rise to vestigial orders such as charge-density waves, loop currents, or charge-$4e$ superconductivity~\cite{PDW,Hamidian2016,WangNP18,JCDavis19,WangPRX21}; for an overview see Refs.~\cite{PDW,RMP04}. Finite-momentum pairing was originally proposed by Fulde and Ferrell (FF)~\cite{FF_state} and by Larkin and Ovchinnikov (LO)~\cite{LO_state} for superconductors in a strong Zeeman field. Recently, an \textit{orbital} analogue---the orbital-FFLO state---has been proposed in engineered heterostructures without large global spin splitting, including bilayer transition-metal dichalcogenides~\cite{bilayerTMD,Lu,Yu,Xi} and Bi$_2$Te$_3$/NbSe$_2$~\cite{Zheng21}. Within a Lawrence--Doniach framework~\cite{LawrenceDoniach,YangKun}, this mechanism was shown~\cite{Qiu2022} to admit both an FF state and a structurally modulated ``Bloch'' SC state.

Very recently, an intrinsic zero-field PDW was discovered in the stoichiometric iron-pnictide EuRbFe$_4$As$_4$~\cite{PDW1144}. Spectroscopic-imaging STM revealed a unidirectional, incommensurate modulation of the SC gap with a wavelength of approximately 8 unit cells. This modulation appears strictly below the magnetic transition temperature $T_m \approx 15\,$K, while uniform superconductivity persists up to $T_c \approx 37\,$K~\cite{Cao2016,Kawashima2016,liu2021iron}. Structurally, EuRbFe$_4$As$_4$ consists of alternating Rb$^+$ and Eu$^{2+}$ planes sandwiching the superconducting FeAs layers. Neutron diffraction identifies a helical Eu spin order with propagation vector $\mathbf{Q} = (0,0,1/4)$, corresponding to a $90^\circ$ rotation of ferromagnetic moments between successive magnetic layers~\cite{Iida2019}. 

The purely electromagnetic orbital-FFLO mechanism~\cite{bilayerTMD,Qiu2022} cannot explain this observation: the dipolar field of the ordered Eu moments ($\mu_0 M\sim 0.4\,$T~\cite{liu2021iron}) yields a modulation length $\sim 5\,\mu$m, three orders of magnitude larger than the observed $\sim 3\,$nm. We propose instead a mechanism rooted in the crystallographic structure of EuRbFe$_4$As$_4$~\cite{liu2021iron}. Each FeAs layer is asymmetrically sandwiched between a non-magnetic Rb$^+$ layer and a magnetic Eu$^{2+}$ layer, which breaks local inversion symmetry and induces a finite Rashba spin--orbit coupling (SOC)~\cite{smidman2017rashba}. The ordered Eu$^{2+}$ moments exert a layer-dependent in-plane exchange field $\mathbf{H}_l$ on the conduction electrons; combined with the Rashba SOC, this generates a layer-dependent Cooper-pair momentum shift, equivalently a formal effective $U(1)$ gauge field $\mathbf{A}_l^{\text{eff}}\propto -\hat{z} \times \mathbf{H}_{l}$~\cite{kaur2005helical, agterberg2012}. 

Because the Eu$^{2+}$ spins rotate by $90^\circ$ between adjacent layers, the induced gauge field has a period-four helical structure along the $c$-axis. We formulate a Ginzburg--Landau theory for the resulting periodically-driven layered superconductor and show that this Rashba--exchange mechanism naturally stabilizes a uniaxial PDW at the observed nanometer scale, with characteristic interlayer loop-current signatures.
	
	\begin{figure}[tb]
		\begin{center}
			
			\subfigure[]{
				\label{fig1a}
				\begin{tikzpicture}[scale=0.6, every node/.style={scale=0.85}]
					\def\dy{1.6} 
					
					\draw[-{Latex[length=2mm]}, thick] (3.2, 0.0) -- (4.2, 0.0) node[above] {$x$};
					\draw[-{Latex[length=2mm]}, thick] (3.2, 0.0) -- (3.8, 0.6) node[above right, inner sep=1pt] {$y$};
					\draw[-{Latex[length=2mm]}, thick] (3.2, 0.0) -- (3.2, 1.0) node[above] {$z$};
					
					\foreach \l in {0,1,2,3,4} {
						
						\filldraw[fill=blue!15, draw=blue!50!black, thick, rounded corners=0.5mm] 
						(-2, \l*\dy) -- (2, \l*\dy) -- (3, \l*\dy+1) -- (-1, \l*\dy+1) -- cycle;
						
						\node[left] at (-2.2, \l*\dy+0.5) {\large $l=\l$};
						
						\ifnum\l<4
						\draw[gray, thick, decoration={coil, aspect=0.3, segment length=2mm, amplitude=0.8mm}, decorate] 
						(0.5, \l*\dy+0.5) -- (0.5, \l*\dy+\dy+0.5);
						\fi
						
						\ifnum\l=0
						\node[right, gray] at (0.7, \l*\dy+0.5*\dy+0.5) {\large $g$};
						\fi
						
						\ifnum\l=0
						\draw[thick, blue!60!black] 
						(-1.5, \l*\dy+0.25) 
						sin (-1.0, \l*\dy+0.45) cos (-0.5, \l*\dy+0.25) 
						sin (0.0, \l*\dy+0.05) cos (0.5, \l*\dy+0.25) 
						sin (1.0, \l*\dy+0.45) cos (1.5, \l*\dy+0.25)
						sin (2.0, \l*\dy+0.05) cos (2.5, \l*\dy+0.25);
						\node[above, blue!60!black] at (2.0, \l*\dy+0.25) {$\psi_l(\mathbf{r})$};
						\fi
						
						\ifnum\l=0 
						\draw[-{Latex[length=3mm,width=2.5mm]}, line width=1.5pt, magenta!90!black] (0.5, \l*\dy+0.5) -- ++(-0.7, -0.7);
						\fi
						\ifnum\l=1 
						\draw[-{Latex[length=3mm,width=2.5mm]}, line width=1.5pt, magenta!90!black] (0.5, \l*\dy+0.5) -- ++(1.1, 0) node[above, xshift=2mm] {\large $\mathbf{A}_l^{\text{eff}}$};
						\fi
						\ifnum\l=2 
						\draw[-{Latex[length=3mm,width=2.5mm]}, line width=1.5pt, magenta!90!black] (0.5, \l*\dy+0.5) -- ++(0.7, 0.7);
						\fi
						\ifnum\l=3 
						\draw[-{Latex[length=3mm,width=2.5mm]}, line width=1.5pt, magenta!90!black] (0.5, \l*\dy+0.5) -- ++(-1.1, 0);
						\fi
						\ifnum\l=4 
						\draw[-{Latex[length=3mm,width=2.5mm]}, line width=1.5pt, magenta!90!black] (0.5, \l*\dy+0.5) -- ++(-0.7, -0.7);
						\fi
					}
					
					\draw[|<->|, thick] (3.5, 1*\dy+0.5) -- (3.5, 2*\dy+0.5) node[midway, right] {\large $a$};
				\end{tikzpicture}
			}
			\hfill
			\subfigure[]{
				\label{fig1b}
				\begin{tikzpicture}[scale=0.75, every node/.style={scale=0.9}]
					\def\q{0.5} 
					
					\draw[dashed, gray, thick] (-2.2, 0) -- (2.2, 0) node[below left, black] {\large $k_x$};
					\draw[dashed, gray, thick] (0, -1.8) -- (0, 2.5) node[below right, black] {\large $k_y$};
					
					\draw[line width=1.2pt, cyan!80!blue] (0, \q) circle (0.9);
					\draw[line width=1.2pt, blue!80!black] (0, \q) circle (1.5);
					
					\fill (0,0) circle (2pt);
					\fill (0,\q) circle (2pt);
					
					\draw[-{Latex[length=2.5mm,width=2mm]}, line width=1.5pt, magenta!90!black] 
					(0,\q-0.05) -- (0,0) node[midway, right] {\large $\mathbf{q}$};
					
					\draw[-{Latex[length=2.5mm,width=2mm]}, line width=2pt, green!60!black] 
					(0, -1.3) -- (1.2, -1.3) node[right] {\large $\mathbf{H}_{l}$};
				\end{tikzpicture}
			}
		\caption{(a) Schematic of the period-4 layered superconductor model. Blue slabs represent the superconducting layers ($l=0,1,2,3,4$) coupled by Josephson tunneling $g$. Bold magenta arrows indicate the effective gauge field $\mathbf{A}_l^{\text{eff}}$, rotating by $90^\circ$ per layer. The sinusoidal curve illustrates the spatially modulated PDW order parameter. (b) Microscopic mechanism generating the effective gauge field. In the presence of Rashba spin-orbit coupling, an in-plane exchange field $\mathbf{H}_{l}$ asymmetrically shifts the Fermi surfaces in momentum space ($k_x, k_y$). This imparts a finite center-of-mass momentum $\mathbf{q} \propto -\hat{z} \times \mathbf{H}_{l}$ to the Cooper pairs in the convention used below, acting strictly as an effective $U(1)$ gauge field orthogonal to the local magnetic moment.
				\label{fig1}}
		\end{center}
	\end{figure}
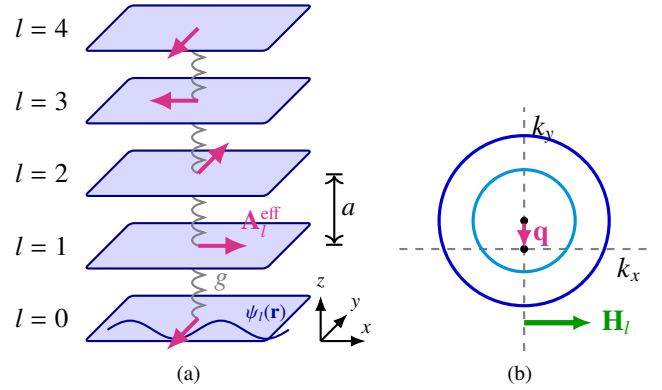
	
\textbf{Model.} We consider a layered superconductor where the basic SC unit cell consists of four effective two-dimensional (2D) SC layers, indexed by $l=0,1,2,3$, aligned along the $c$-axis (which we set as the $z$-axis). In EuRbFe$_4$As$_4$, the asymmetric chemical environment of the FeAs planes (sandwiched between Rb$^+$ and Eu$^{2+}$) breaks local inversion symmetry, inducing a Rashba spin-orbit coupling with the Rashba vector $\boldsymbol{\alpha}_R \parallel \hat{z}$. Concurrently, the helical magnetic order of the Eu$^{2+}$ ions generates a localized, in-plane exchange field $\mathbf{H}_l$ acting on the conduction electrons of the $l$-th SC layer. 

The $\mathbf{Q} = (0,0,1/4)$ helical Eu order produces a layer-dependent exchange field on the conduction electrons of the $l$-th SC layer (see Supplemental Material for the construction):
\begin{equation}
	\mathbf{H}_l = H_{\text{ex}} \left( \cos\frac{l\pi}{2}\hat{x} + \sin\frac{l\pi}{2}\hat{y} \right),
\end{equation}
where $H_{\text{ex}}$ is the exchange-energy scale entering the electronic Hamiltonian. The combination of Rashba SOC and this in-plane exchange field shifts the Fermi surfaces asymmetrically, imparting a finite center-of-mass momentum $\mathbf{q}_l = k_0\hat{n}_l$ to the Cooper pairs~\cite{kaur2005helical, agterberg2012}, with
\begin{equation}
	\hat{\mathbf{n}}_l = \sin\frac{l\pi}{2}\hat{x}-\cos\frac{l\pi}{2}\hat{y}, \qquad k_0\approx \frac{2|\alpha_R|H_{\text{ex}}}{\hbar^2 v_F^2}>0.
\end{equation}

In the GL framework $\mathbf{q}_l$ enters the covariant derivative $\mathbf{D}_l = \nabla - i\mathbf{q}_l$, mathematically equivalent to a layer-dependent $U(1)$ gauge field $\mathbf{A}_l \equiv (\hbar c/2e) \mathbf{q}_l$:
\begin{equation}\label{eq:Al}
	\mathbf{A}_l = H_{\text{eff}}\, a \left(\sin\frac{l\pi}{2}\hat{x}-\cos\frac{l\pi}{2}\hat{y}\right), \quad H_{\text{eff}}\equiv\frac{\hbar c\,k_0}{2ea}>0.
\end{equation}
Here $\mathbf{A}_l$ is \textit{not} a physical electromagnetic vector potential and produces no orbital pair-breaking; it is a formal representation of the Rashba--exchange momentum shift. Using $\alpha_R\sim 0.1\,$eV\,\AA, $H_{\text{ex}} \sim 26\,$meV, $v_F \sim 4 \times 10^4\,$m/s~\cite{liu2021iron} yields a nanometer-scale modulation $\lambda=2\pi/k_0\sim3\,$nm (see Supplemental Material), versus the micron scale expected from the bare Eu dipole field. The gauge field satisfies $\mathbf{A}_{l+4}=\mathbf{A}_l$, encoding the period-four helical structure [Fig.~\ref{fig1}(a)].
	
\textbf{Symmetry and SC Order Parameter.} The zone-center hole pockets in EuRbFe$_4$As$_4$ are predominantly $3d_{xz}/3d_{yz}$ in character~\cite{Kim2021, Hemmida2021} and form a 2D irreducible representation $\Gamma_5$ of the crystal point group $D_4$. We therefore work with a two-component orbital order parameter $\vec{\psi}_l \equiv (\psi_{l,xz}, \psi_{l,yz})^T$ on each layer. The gauge field $\mathbf{A}_l$ also transforms as $\Gamma_5$, breaks the local $C_4$ symmetry, and asymmetrically shifts the $d_{xz}/d_{yz}$ Fermi surfaces, suppressing the $\mathbf{q}=0$ pairing channel and selecting an orbital-selective finite-momentum state. The magnetic space group $P_{c}4_{1}22$ further constrains the layer dependence: the $4_1$ screw axis enforces $\vec{\psi}_{l+1}(-y,x) = -i\sigma_y \vec{\psi}_l(x,y)$, where $\sigma_i$ are Pauli matrices on the orbital basis (full analysis in Supplemental Material).

\begin{figure*}[tb]
	\begin{center}
		\subfigure[Diagonal Nematic]{\includegraphics[width=0.32\linewidth]{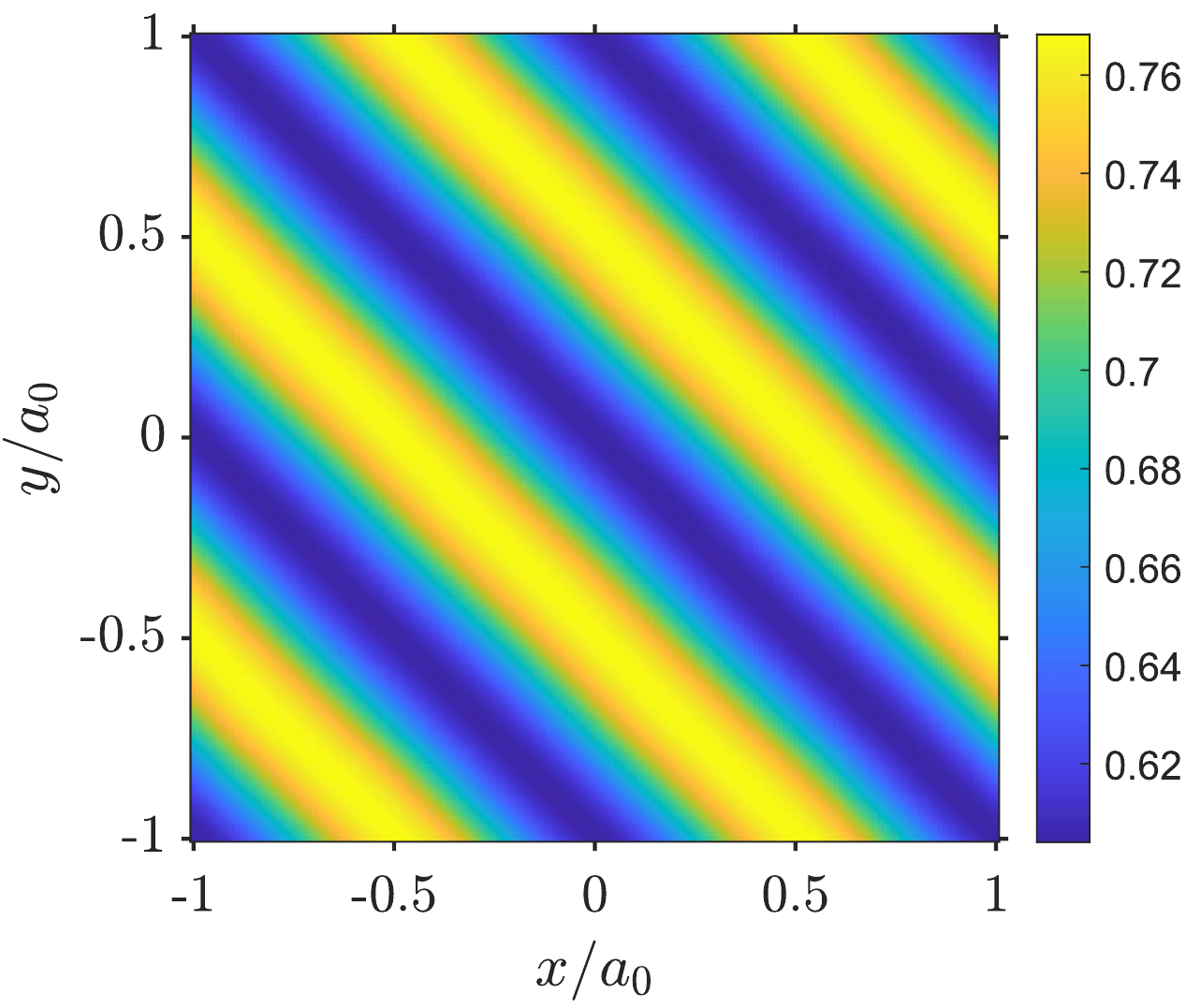}}\hfill
		\subfigure[Principal Nematic]{\includegraphics[width=0.32\linewidth]{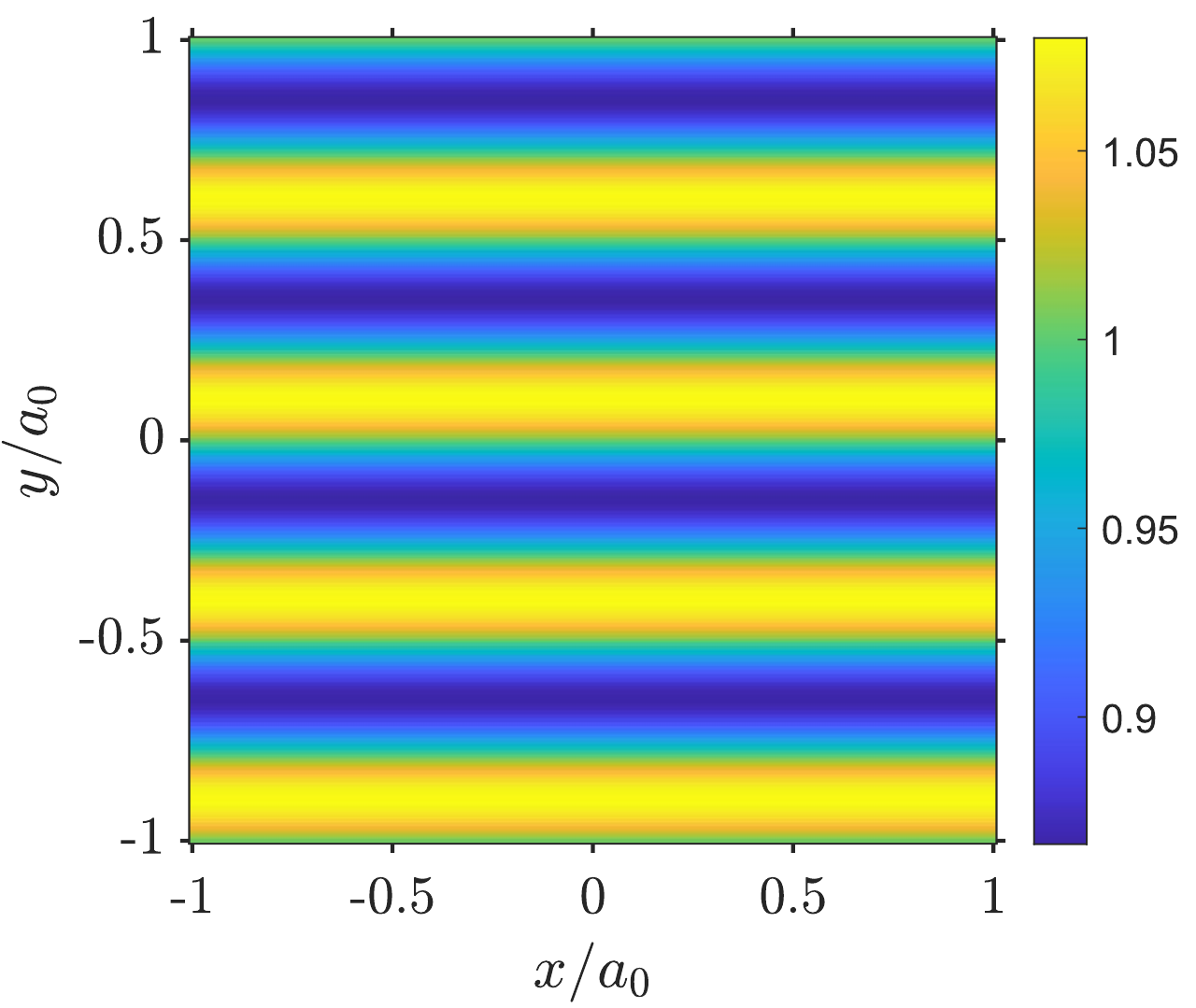}}\hfill
		\subfigure[Chiral]{\includegraphics[width=0.32\linewidth]{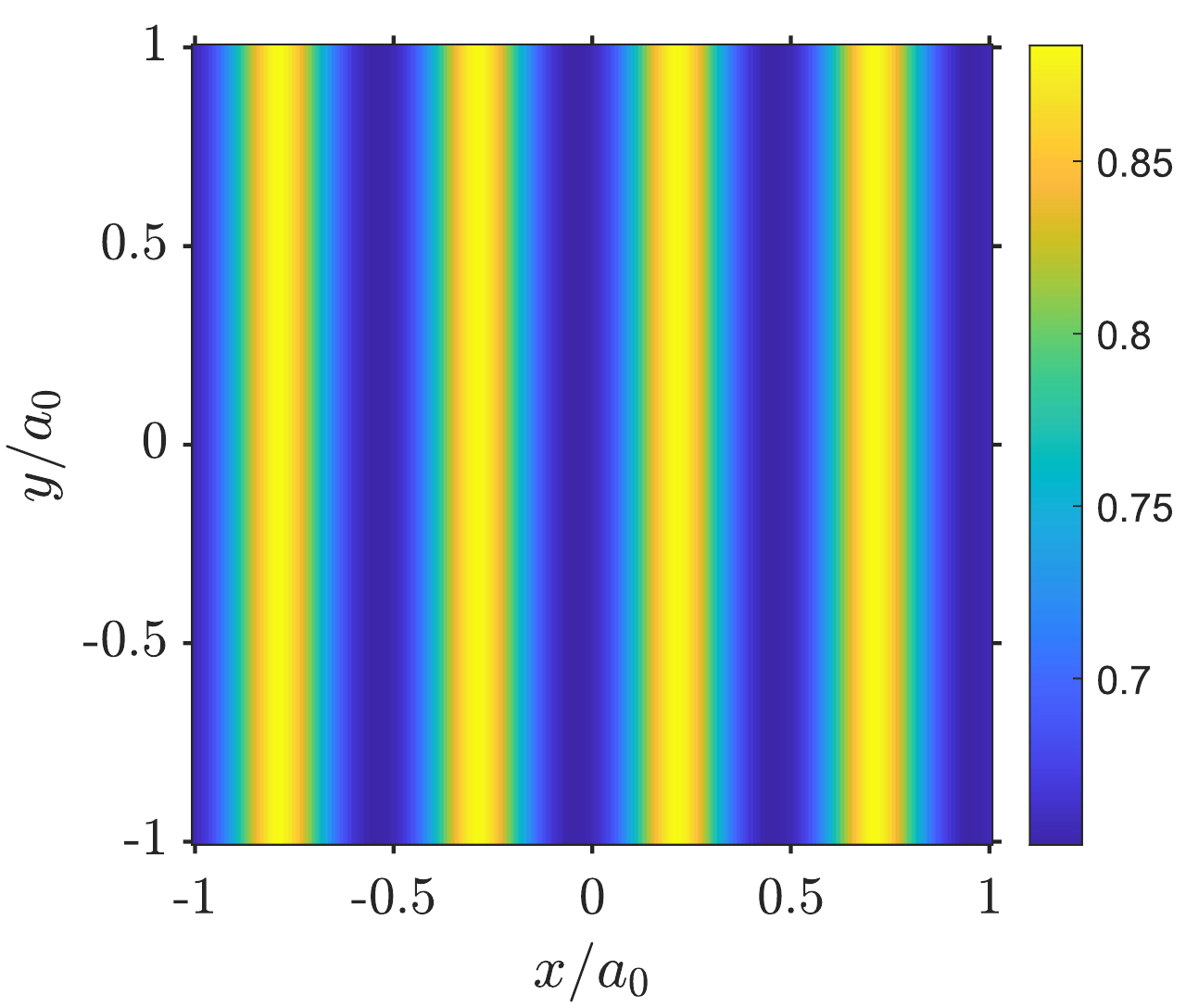}}
		\caption{Superfluid density distributions of the Bloch SC states on layer $l=0$ with $a_0=2\pi/k_0$. In all cases, the system spontaneously breaks rotational symmetry, resulting in a strictly uniaxial (unidirectional) PDW modulation, consistent with STM observations. Here $\beta_1$ and $\beta_{2,3}$ are the quartic coefficients of Eq.~\eqref{eq:free2}; the dimensionless effective field is $h^2 \equiv \hbar^2 k_0^2/(2m^*|\alpha|)$. Common parameters: $\beta_1=3$, $h=2.5$. Specific parameters: (a) Diagonal nematic ($\beta_2=1, \beta_3=1, g_0=1, g_2=0$); (b) Principal nematic ($\beta_2=1, \beta_3=-1, g_0=0, g_2=1$); (c) Chiral ($\beta_2=-1, \beta_3=1, g_0=0, g_2=1$).}
		\label{fig:density}
	\end{center}
\end{figure*}
	
\textbf{GL formulation.} The GL free-energy density is built from $\vec{\psi}_l$ and $\mathbf{D}_l = \nabla - i k_0 \hat{n}_l$, where $k_0\hat{n}_l = (2\pi/\Phi_0)\mathbf{A}_l = \mathbf{q}_l$ and $\Phi_0 = hc/2e$. We adopt a $D_4$-symmetric soft-mode gradient form (the first term in Eq.~\eqref{eq:free2}) which exposes the uniaxial PDW instability transparently; generic $D_4$ stiffnesses lift exact zero modes and shift phase boundaries but preserve the symmetry-allowed finite-momentum tendency (full classification in Supplemental Material). The free energy invariant under $P_{c}4_{1}22$ reads:
\begin{equation}\label{eq:free2}
	\begin{split}
		&f_s - f_n = \sum_{l=0}^{3} \left[ \frac{\hbar^2}{2m^*} \left( \left|\mathbf{D}_l\cdot\vec{\psi}_l\right|^2+\left|\mathbf{D}_l\cdot\sigma_x\vec{\psi}_l\right|^2 \right) + \alpha \vec{\psi}_l^\dagger \vec{\psi}_l \right. \\
		& \quad - g_0 \left( \vec{\psi}_l^\dagger \vec{\psi}_{l+1} + \text{c.c.} \right) - g_2 \left( \vec{\psi}_l^\dagger (i\sigma_y)\vec{\psi}_{l+1} + \text{c.c.} \right) \\
		& \left. \quad + \frac{\beta_1}{2} \left( \vec{\psi}_l^\dagger \vec{\psi}_l \right)^2  + \frac{\beta_2}{2} \left( \vec{\psi}_l^\dagger \sigma_y \vec{\psi}_l \right)^2 + \frac{\beta_3}{2} \left( \vec{\psi}_l^\dagger \sigma_z \vec{\psi}_l \right)^2 \right],
	\end{split}
\end{equation}
where $\mathbf{D}_l\cdot\vec{\psi}_l \equiv D_{lx}\psi_{l,1}+D_{ly}\psi_{l,2}$, $\mathbf{D}_l\cdot\sigma_x\vec{\psi}_l \equiv D_{lx}\psi_{l,2}+D_{ly}\psi_{l,1}$, layer indices are mod 4, and stability requires $\beta_1 > 0$, $\beta_{2} > -\beta_1$, and $\beta_{3} > -\beta_1$.
	
	\textbf{Decoupled and Bloch SC states.} In the absence of inter-layer coupling ($g_0 = g_2 = 0$), the system admits a uniform plane-wave solution $\vec{\psi}_l(\mathbf{r}) = \sqrt{\rho_s} ( u_l, v_l )^T e^{i k_0 \hat{n}_l \cdot \mathbf{r}}$, where $\rho_s$ is the uniform superfluid density. Depending on the quartic parameters $\beta_2$ and $\beta_3$, the spinor aligns into one of three decoupled ground states:
\begin{equation}
	(u_l, v_l) = 
	\begin{cases} 
		\frac{1}{\sqrt{2}} (1, \pm 1), & \beta_2 > 0, \, \beta_3 > 0, \\
		(1, 0) \text{ or } (0, 1), & \beta_3 < 0, \, \beta_2 > \beta_3, \\
		\frac{1}{\sqrt{2}} (1, \pm i), & \beta_2 < 0, \, \beta_3 > \beta_2.
	\end{cases}
\end{equation}
These correspond, respectively, to \textit{diagonal nematic}, \textit{principal nematic}, and \textit{chiral} (time-reversal-breaking) states.

Upon introducing Josephson coupling ($g_0, g_2 \neq 0$), the competition between phase locking and the layer-dependent effective gauge field drives a spatial modulation of the order parameter. We adopt a Bloch wave-function ansatz:
\begin{equation}\label{eq:BlochSC2}
	\vec{\psi}_l(\mathbf{r}) = \vec{\varphi}_l(\mathbf{r})\, e^{ik_{0}\hat{n}_{l}\cdot\mathbf{r}},
\end{equation}
where the 2D periodic envelope $\vec{\varphi}_l$ is expanded in Fourier harmonics. A perturbative analysis (see Supplemental Material) reveals that the rotating gauge field generates an exact zero-energy soft mode in the diagonal spatial gradients. Depending on the specific decoupled ground state, the inter-layer Josephson couplings act as symmetry-breaking sources that project directly onto this soft mode. This geometric coupling destabilizes the decoupled plane-wave state, driving the formation of the spatially modulated Bloch SC state. While generic $D_4$-invariant gradient coefficients lift these perfect zero modes into finite-energy soft modes, the system still robustly falls into a 1D uniaxial (stripe-like) Bloch state, merely shifting the critical Josephson threshold.

Numerical minimization yields a sequence: decoupled $\to$ FF $\to$ Bloch SC state as the effective field increases at fixed Josephson coupling (full phase diagrams in Supplemental Material).
	
\begin{figure*}[tb]
	\begin{center}
		\subfigure[Diagonal Nematic]{\includegraphics[width=0.33\linewidth]{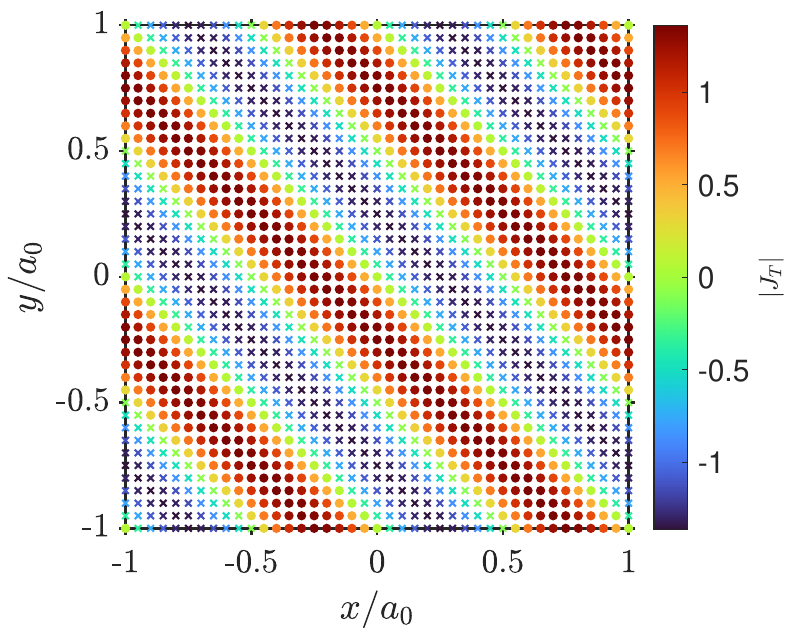}}\hfill
		\subfigure[Principal Nematic]{\includegraphics[width=0.33\linewidth]{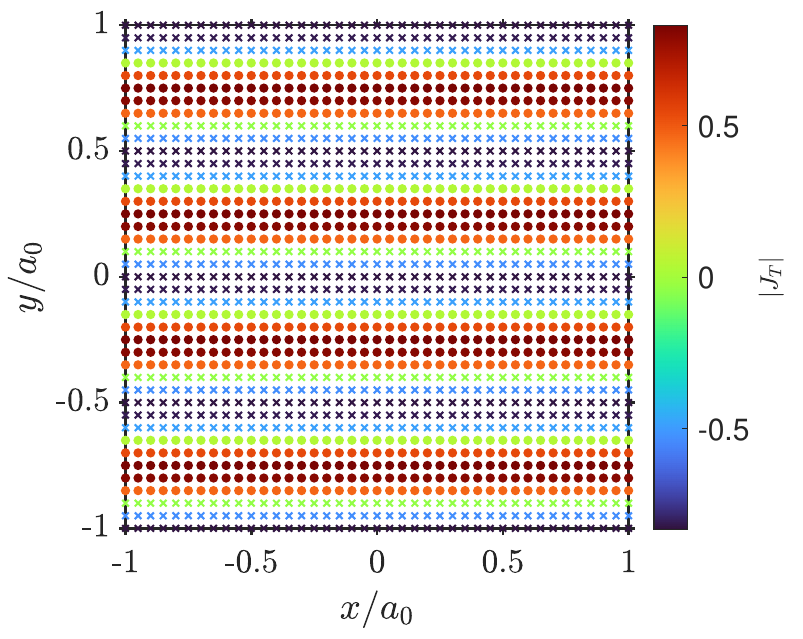}}\hfill
		\subfigure[Chiral]{\includegraphics[width=0.33\linewidth]{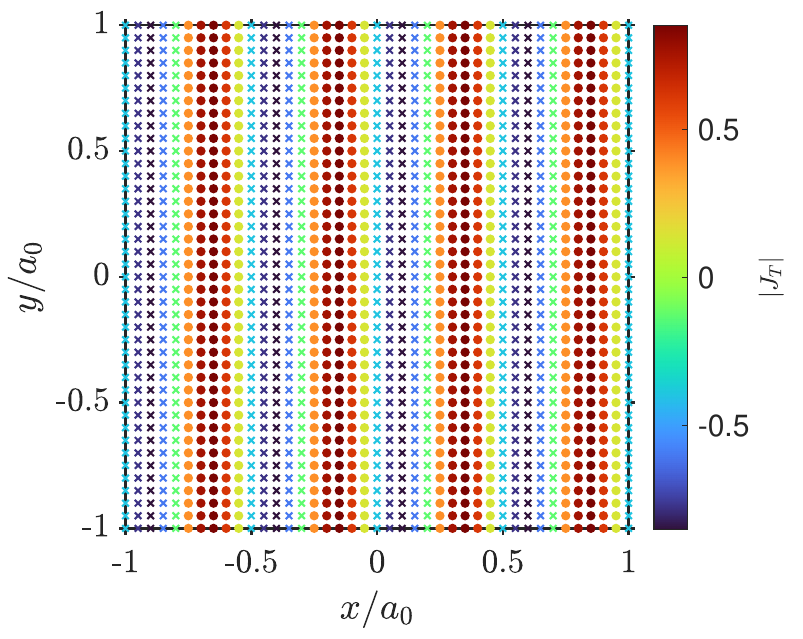}}
		\caption{Inter-layer Josephson tunneling currents between layers $l=0$ and $l=1$ for the corresponding Bloch SC states shown in Fig.~\ref{fig:density}. The highly structured, uniaxial modulation of the inter-layer phase coherence is a direct consequence of the layer-dependent effective gauge field. Dots and crosses in figures denote Josephson currents flowing upward ($+z$) and downward ($-z$), respectively. Parameters for panels (a)--(c) are identical to those in Fig.~\ref{fig:density}(a)--(c), respectively.}
		\label{fig:josephson}
	\end{center}
\end{figure*}

\begin{figure}[tb]
	\begin{center}
		\includegraphics[width=0.95\linewidth]{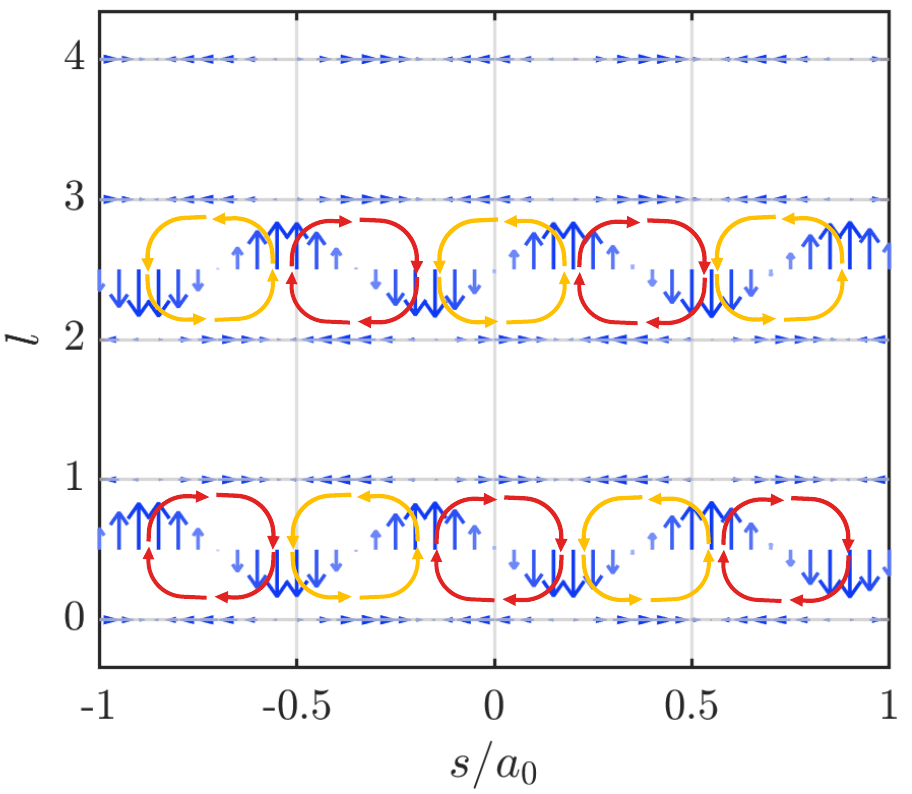}
		\caption{Side-view illustration of the 3D spontaneous currents in the structurally modulated Bloch SC state, viewed along the in-plane $135^\circ$ direction [$(-\hat{x}+\hat{y})/\sqrt{2}$], with the horizontal coordinate $s=(x+y)/\sqrt{2}$. Here, the diagonal nematic state is depicted, which produces density modulations along the diagonal axes (corresponding to the Fe-Fe bond directions in the crystal), in excellent agreement with STM observations~\cite{PDW1144}. The spatially oscillating inter-layer phase difference drives alternating Josephson tunneling currents, forming a periodic array of Josephson vortex-antivortex pairs (red and yellow arrows) across the insulating spacer layers. This circulating out-of-plane transport dynamically couples the intra-layer stripe textures.}
		\label{fig:vortex}
	\end{center}
\end{figure}
	
We illustrate the spatially modulated Bloch SC states for three representative parameter sets, corresponding to the three distinct decoupled ground states (Diagonal Nematic, Principal Nematic, and Chiral). Figure~\ref{fig:density} displays the spatial distribution of the superfluid density, $\vec{\psi}_l^\dagger \vec{\psi}_l = |\psi_{l,1}|^2 + |\psi_{l,2}|^2$, for the $l=0$ layer. Remarkably, regardless of the specific $\beta$ parameters, the system always develops a strictly uniaxial (unidirectional) stripe-like modulation. The modulation direction aligns along the diagonal for the diagonal nematic state [Fig.~\ref{fig:density}(a)] and along the principal axes for the principal nematic and chiral states [Fig.~\ref{fig:density}(b,c)]. This robust uniaxial structure---distinct from the checkerboard patterns produced by scalar 1D representations---captures the unidirectional STM modulation reported in Ref.~\cite{PDW1144}.

To visualize how a unidirectional pattern propagates through the 3D crystal, one must consider the layer-by-layer symmetries. Although the underlying exchange field rotates by $90^\circ$ along the $c$-axis, the two stripe orientations related by a $90^\circ$ rotation are degenerate. The ground state therefore spontaneously selects one common orientation for the uniaxial stripes on all superconducting layers.
The most physically relevant scenario for EuRbFe$_4$As$_4$ is the diagonal nematic state: because the Eu$^{2+}$ spins order along the Fe-As bonds, the gauge field axes ($x, y$) align with the Fe-As directions. The diagonal nematic state thus naturally produces a modulation along the Fe-Fe bonds, in excellent agreement with the experimental STM observations~\cite{PDW1144}. In this state, all layers may choose the same diagonal modulation wave vector, for example along the $x+y$ direction in Fig.~\ref{fig:density}(a), while the $x-y$ direction is its degenerate partner; the real-space stripes run perpendicular to the selected wave vector. This is dynamically coupled by the inter-layer Josephson tunneling currents (Fig.~\ref{fig:josephson}). Viewed from the side, the spatially oscillating inter-layer phase difference forces the out-of-plane transport to periodically reverse direction, generating a dense, regular array of Josephson vortex-antivortex pairs across the spacer layers (Fig.~\ref{fig:vortex}). The inter-layer current distribution also develops a characteristic dimer-like pattern along the side-view direction, with neighboring current loops grouped into paired structures.
	
	\textbf{Discussion and Conclusion.}  
Our phenomenological Ginzburg-Landau analysis demonstrates that the combination of local inversion symmetry breaking (Rashba SOC) and a helical magnetic exchange field naturally generates a layer-dependent effective $U(1)$ gauge field. When applied to the period-four magnetic structure of EuRbFe$_4$As$_4$, this effective gauge field naturally stabilizes uniaxial pair density wave states. 

The condensation into the 2D representation $\Gamma_5$ is microscopically mandated by the interplay of orbital physics and magnetic symmetry breaking. Between $T_c \approx 37\,$K and $T_m \approx 15\,$K, uniform superconductivity dominates. However, when the helical magnetic order sets in below $T_m$, the induced effective gauge field asymmetrically shifts the degenerate Fe $3d_{xz}/3d_{yz}$ hole pockets. Because this gauge field explicitly breaks local rotational symmetry, it suppresses the uniform $q=0$ state in these bands and forces the Cooper pairs to acquire a finite momentum. This provides a rigorous symmetry-based explanation for the STM observation that the unidirectional PDW is not an independent instability, but emerges precisely below $T_m$ as a consequence of the helical magnetic order.

Quantitatively, the dimensionless momentum shift can be estimated as $q/k_F \approx (|\alpha_R| k_F) H_{\text{ex}} / (2 E_F^2)$. Iron pnictides are characterized by notably shallow Fermi pockets; for the relevant $d_{xz}/d_{yz}$ bands, $E_F \sim 10$--$15\,$meV and $k_F \sim 0.10$--$0.12\,\text{\AA}^{-1}$. The exchange field $H_{\text{ex}}$ on the conduction electrons, estimated via the RKKY mechanism from the $15\,$K ordering temperature, is $H_{\rm ex} \approx 26\,$meV. Combined with a typical Rashba spin splitting $|\alpha_R| k_F \sim 10$--$20\,$meV, the perturbative expression gives $q/k_F \sim 1$--$2$, corresponding to a PDW wavelength $\lambda = 2\pi/q$ of a few nanometers, comparable to the $\sim 8$-unit-cell STM modulation~\cite{PDW1144}. Since this estimate lies at the edge of the controlled weak-exchange regime, it should be regarded as an order-of-magnitude indication that the Rashba-exchange mechanism naturally reaches the nanometer scale; a fully quantitative wavelength requires a non-perturbative multiband band-structure calculation.

EuRbFe$_4$As$_4$ is a multiband superconductor, and the present GL theory focuses on the shallow $d_{xz}/d_{yz}$ hole pockets because they have the correct $\Gamma_5$ orbital symmetry and the smallest Fermi energies, making them the most susceptible to a Rashba-exchange momentum shift. Other hole and electron pockets can provide additional uniform superconducting stiffness and interband Josephson locking, and may renormalize the critical fields and modulation wave vector. Phenomenologically, such effects are absorbed into the coefficients $\alpha$, $g_0$, $g_2$, and $\beta_i$ in Eq.~\eqref{eq:free2}; the GL theory commits only to the $\Gamma_5$ orbital symmetry of the dominant finite-momentum channel, not to a single-band microscopic model. These additional bands do not, however, remove the symmetry-allowed finite-momentum tendency in the $d_{xz}/d_{yz}$ sector; a microscopic multiband treatment is an important extension beyond the scope of the present phenomenological theory.

Crucially, our phase diagrams (see Supplemental Material) reveal that the structurally modulated Bloch SC state is stabilized in the regime where the inter-layer Josephson coupling $g$ is relatively weak compared to the effective gauge field energy. This is highly consistent with the crystallographic nature of EuRbFe$_4$As$_4$. Because the superconducting FeAs planes are separated by alternating, chemically distinct Rb$^+$ and Eu$^{2+}$ spacer layers, the material is a highly anisotropic quasi-2D superconductor with a notably short $c$-axis coherence length~\cite{liu2021iron}. Consequently, the inter-layer coupling is intrinsically weak, placing EuRbFe$_4$As$_4$ squarely within the parameter regime where our theory predicts the uniaxial PDW to be the robust ground state.

Furthermore, our theory yields a characteristic prediction: the uniaxial PDW states are accompanied by complex, spontaneously generated 3D current patterns. The intra-layer uniaxial flow patterns (see Supplemental Material) are coupled by the modulated inter-layer Josephson tunneling (Fig.~\ref{fig:josephson}), which manifests macroscopically as a periodic lattice of Josephson vortex-antivortex pairs (Fig.~\ref{fig:vortex}). We estimate the local internal fields generated by these vestigial 3D currents to be of order $\sim 0.1$--$1\,$mT, oscillating at the PDW wave vector. Although small compared with the uniform Eu dipolar background ($\mu_0 M\sim 0.4\,$T), the loop-current signal lives at finite $\mathbf{q}$ and is in principle separable by zero-field muon spin relaxation ($\mu$SR) or scanning SQUID microscopy. Such a measurement would be a definitive test of the helical gauge-field mechanism in EuRbFe$_4$As$_4$.
	
\begin{acknowledgments}
	We thank Guang-Han Cao for helpful discussions. This work was supported by the National Key R\&D Program of China (Grant No.\ 2022YFA1403403), the National Natural Science Foundation of China (Grants No.\ 12274441 and 12534004), and a fellowship award from the Hong Kong Research Grant Council (Project No.\ SRFS2324-6S01).
\end{acknowledgments}

\bibliographystyle{apsrev4-1}
\bibliography{reference}

@article{YXWang-PRB2018,
title = {Pair density waves in superconducting vortex halos},
author = {Wang, Yuxuan and Edkins, Stephen D. and Hamidian, Mohammad H. and Davis, J. C. S\'eamus and Fradkin, Eduardo and Kivelson, Steven A.},
journal = {Phys. Rev. B},
volume = {97},
issue = {17},
pages = {174510},
numpages = {13},
year = {2018},
month = {May},
publisher = {American Physical Society},
doi = {10.1103/PhysRevB.97.174510},
url = {https://link.aps.org/doi/10.1103/PhysRevB.97.174510}
}

@article{LFu-PRB2020,
title = {Charge transfer excitations, pair density waves, and superconductivity in moir\'e materials},
author = {Slagle, Kevin and Fu, Liang},
journal = {Phys. Rev. B},
volume = {102},
issue = {23},
pages = {235423},
numpages = {11},
year = {2020},
month = {Dec},
publisher = {American Physical Society},
doi = {10.1103/PhysRevB.102.235423},
url = {https://link.aps.org/doi/10.1103/PhysRevB.102.235423}
}

@article{Yao-arX2022,
title = {Pair-density-wave and chiral superconductivity in twisted bilayer transition-metal-dichalcogenides},
author = {Wu, Yi-Ming and Wu, Zhengzhi and Yao, Hong},
journal = {Phys. Rev. Lett.},
volume = {130},
issue = {12},
pages = {126001},
year = {2023},
doi = {10.1103/PhysRevLett.130.126001}
}

@article{Kim-SciAdv2019,
title = {Evidence of pair-density wave in spin-valley locked systems},
author = {Venderley, Jordan and Kim, Eun-Ah},
journal = {Sci. Adv.},
volume = {5},
number = {3},
pages = {eaat4698},
year = {2019},
publisher = {American Association for the Advancement of Science}
,
doi = {10.1126/sciadv.aat4698}
,
url = {https://doi.org/10.1126/sciadv.aat4698}
}

@article{Fradkin-PRB2012,
title = {Pair-density-wave superconducting order in two-leg ladders},
author = {Jaefari, Akbar and Fradkin, Eduardo},
journal = {Phys. Rev. B},
volume = {85},
issue = {3},
pages = {035104},
numpages = {19},
year = {2012},
month = {Jan},
publisher = {American Physical Society},
doi = {10.1103/PhysRevB.85.035104},
url = {https://link.aps.org/doi/10.1103/PhysRevB.85.035104}
}

@article{YXWang-PRL2015,
title = {Coexistence of Charge-Density-Wave and Pair-Density-Wave Orders in Underdoped Cuprates},
author = {Wang, Yuxuan and Agterberg, Daniel F. and Chubukov, Andrey},
journal = {Phys. Rev. Lett.},
volume = {114},
issue = {19},
pages = {197001},
numpages = {6},
year = {2015},
month = {May},
publisher = {American Physical Society},
doi = {10.1103/PhysRevLett.114.197001},
url = {https://link.aps.org/doi/10.1103/PhysRevLett.114.197001}
}

@article{Berg-PRL2010,
title = {Pair-Density-Wave Correlations in the Kondo-Heisenberg Model},
author = {Berg, Erez and Fradkin, Eduardo and Kivelson, Steven A.},
journal = {Phys. Rev. Lett.},
volume = {105},
issue = {14},
pages = {146403},
numpages = {4},
year = {2010},
month = {Sep},
publisher = {American Physical Society},
doi = {10.1103/PhysRevLett.105.146403},
url = {https://link.aps.org/doi/10.1103/PhysRevLett.105.146403}
}

@article{ZYHan-PRL2020,
title = {Strong Coupling Limit of the Holstein-Hubbard Model},
author = {Han, Zhaoyu and Kivelson, Steven A. and Yao, Hong},
journal = {Phys. Rev. Lett.},
volume = {125},
issue = {16},
pages = {167001},
numpages = {6},
year = {2020},
month = {Oct},
publisher = {American Physical Society},
doi = {10.1103/PhysRevLett.125.167001},
url = {https://link.aps.org/doi/10.1103/PhysRevLett.125.167001}
}

@article{Huang-npjQM2022,
title = {Pair-density-wave in the strong coupling limit of the Holstein-Hubbard model},
author = {Huang, Kevin S and Han, Zhaoyu and Kivelson, Steven A and Yao, Hong},
journal = {npj Quantum Mater.},
volume = {7},
number = {1},
pages = {17},
year = {2022},
publisher = {Nature Publishing Group},
url = {https://www.nature.com/articles/s41535-022-00426-w}
}

@article{PDW-Kivelson2007,
title = {Dynamical Layer Decoupling in a Stripe-Ordered High-${T}_{c}$ Superconductor},
author = {Berg, E. and Fradkin, E. and Kim, E.-A. and Kivelson, S. A. and Oganesyan, V. and Tranquada, J. M. and Zhang, S. C.},
journal = {Phys. Rev. Lett.},
volume = {99},
issue = {12},
pages = {127003},
numpages = {4},
year = {2007},
month = {Sep},
publisher = {American Physical Society},
doi = {10.1103/PhysRevLett.99.127003},
url = {https://link.aps.org/doi/10.1103/PhysRevLett.99.127003}
}

@article{Agterberg2008NPdislocations,
title = {Dislocations and vortices in pair-density-wave superconductors},
author = {Agterberg, DF and Tsunetsugu, H},
journal = {Nat. Phys.},
volume = {4},
number = {8},
pages = {639--642},
year = {2008},
publisher = {Nature Publishing Group}
,
doi = {10.1038/nphys984}
,
url = {https://doi.org/10.1038/nphys984}
}

@article{PALee-PRX2014,
title = {Amperean Pairing and the Pseudogap Phase of Cuprate Superconductors},
author = {Lee, Patrick A.},
journal = {Phys. Rev. X},
volume = {4},
issue = {3},
pages = {031017},
numpages = {13},
year = {2014},
month = {Jul},
publisher = {American Physical Society},
doi = {10.1103/PhysRevX.4.031017},
url = {https://link.aps.org/doi/10.1103/PhysRevX.4.031017}
}

@article{maciejko2014weyl,
title = {Weyl semimetals with short-range interactions},
author = {Maciejko, Joseph and Nandkishore, Rahul},
journal = {Phys. Rev. B},
volume = {90},
number = {3},
pages = {035126},
year = {2014},
publisher = {APS}
,
doi = {10.1103/PhysRevB.90.035126}
,
url = {https://doi.org/10.1103/PhysRevB.90.035126}
}

@article{SKJian-PRL2015,
title = {Emergent Spacetime Supersymmetry in 3D Weyl Semimetals and 2D Dirac Semimetals},
author = {Jian, Shao-Kai and Jiang, Yi-Fan and Yao, Hong},
journal = {Phys. Rev. Lett.},
volume = {114},
issue = {23},
pages = {237001},
numpages = {5},
year = {2015},
month = {Jun},
publisher = {American Physical Society},
doi = {10.1103/PhysRevLett.114.237001},
url = {https://link.aps.org/doi/10.1103/PhysRevLett.114.237001}
}

@article{SKJian-PRL2017,
title = {Emergence of Supersymmetric Quantum Electrodynamics},
author = {Jian, Shao-Kai and Lin, Chien-Hung and Maciejko, Joseph and Yao, Hong},
journal = {Phys. Rev. Lett.},
volume = {118},
issue = {16},
pages = {166802},
numpages = {6},
year = {2017},
month = {Apr},
publisher = {American Physical Society},
doi = {10.1103/PhysRevLett.118.166802},
url = {https://link.aps.org/doi/10.1103/PhysRevLett.118.166802}
}

@article{PDW,
title = {The Physics of Pair-Density Waves: Cuprate Superconductors and Beyond},
volume = {11},
doi = {10.1146/annurev-conmatphys-031119-050711},
number = {1},
journal = {Annu. Rev. Condens. Matter Phys.},
publisher = {Annual Reviews},
author = {Agterberg, Daniel F. and Davis, J.C. Séamus and Edkins, Stephen D. and Fradkin, Eduardo and Van Harlingen, Dale J. and Kivelson, Steven A. and Lee, Patrick A. and Radzihovsky, Leo and Tranquada, John M. and Wang, Yuxuan},
year = {2020},
month = {Mar},
pages = {231--270}
}

@article{LO_state,
author = {{Larkin}, A.~I. and {Ovchinnikov}, Yu. N.},
title = "{Quasiclassical Method in the Theory of Superconductivity}",
journal = {Sov. Phys. JETP},
year = {1969},
month = {jun},
volume = {28},
pages = {1200},
url = {https://ui.adsabs.harvard.edu/abs/1969JETP...28.1200L}
}

@article{FF_state,
title = {Superconductivity in a Strong Spin-Exchange Field},
author = {Fulde, Peter and Ferrell, Richard A.},
journal = {Phys. Rev.},
volume = {135},
issue = {3A},
pages = {A550--A563},
numpages = {0},
year = {1964},
month = {Aug},
publisher = {American Physical Society},
doi = {10.1103/PhysRev.135.A550},
url = {https://link.aps.org/doi/10.1103/PhysRev.135.A550}
}

@article{M-F_model_2,
title = {Pair-density-wave superconducting states and electronic liquid-crystal phases},
author = {Soto-Garrido, Rodrigo and Fradkin, Eduardo},
journal = {Phys. Rev. B},
volume = {89},
issue = {16},
pages = {165126},
numpages = {19},
year = {2014},
month = {Apr},
publisher = {American Physical Society},
doi = {10.1103/PhysRevB.89.165126},
url = {https://link.aps.org/doi/10.1103/PhysRevB.89.165126}
}

@article{Berg-cuprate-NJP2009,
year = {2009},
month = {nov},
volume = {11},
number = {11},
pages = {115004},
author = {Erez Berg and Eduardo Fradkin and Steven A Kivelson and John M Tranquada},
title = {Striped superconductors: how spin, charge and superconducting orders intertwine in the cuprates},
publisher = {{IOP} Publishing},
journal = {New J. Phys.},
doi = {10.1088/1367-2630/11/11/115004},
url = {https://doi.org/10.1088/1367-2630/11/11/115004}
}

@article{RMP04,
title = {Inhomogeneous superconductivity in condensed matter and \protect{QCD}},
author = {Casalbuoni, Roberto and Nardulli, Giuseppe},
journal = {Rev. Mod. Phys.},
volume = {76},
issue = {1},
pages = {263--320},
numpages = {0},
year = {2004},
month = {Feb},
publisher = {American Physical Society},
doi = {10.1103/RevModPhys.76.263},
url = {https://link.aps.org/doi/10.1103/RevModPhys.76.263}
}

@article{zhou2021chern,
title = {Chern Fermi pocket, topological pair density wave, and charge-4$e$ and charge-6$e$ superconductivity in kagom\'{e} superconductors},
author = {Zhou, Sen and Wang, Ziqiang},
journal = {Nat. Commun.},
volume = {13},
pages = {7288},
year = {2022},
doi = {10.1038/s41467-022-34832-2}
}

@article{WangPRX21,
title = {Evolution of Charge and Pair Density Modulations in Overdoped \protect{${\mathrm{Bi}}_{2}{\mathrm{Sr}}_{2}{\mathrm{CuO}}_{6+\ensuremath{\delta}}$}},
author = {Li, Xintong and Zou, Changwei and Ding, Ying and Yan, Hongtao and Ye, Shusen and Li, Haiwei and Hao, Zhenqi and Zhao, Lin and Zhou, Xingjiang and Wang, Yayu},
journal = {Phys. Rev. X},
volume = {11},
issue = {1},
pages = {011007},
numpages = {11},
year = {2021},
month = {Jan},
publisher = {American Physical Society},
doi = {10.1103/PhysRevX.11.011007},
url = {https://link.aps.org/doi/10.1103/PhysRevX.11.011007}
}

@article{WangNP18,
title = {Visualization of the periodic modulation of \protect{Cooper} pairing in a cuprate superconductor},
author = {Ruan, Wei and Li, Xintong and Hu, Cheng and Hao, Zhenqi and Li, Haiwei and Cai, Peng and Zhou, Xingjiang and Lee, Dung-Hai and Wang, Yayu},
journal = {Nat. Phys.},
volume = {14},
issue = {12},
pages = {1178--1182},
numpages = {},
year = {2018},
month = {Dec},
doi = {10.1038/s41567-018-0276-8},
url = {https://doi.org/10.1038/s41567-018-0276-8}
}

@article{Hamidian2016,
title = {Detection of a \protect{Cooper-pair} density wave in \protect{Bi$_2$Sr$_2$CaCu$_2$O$_{8+x}$}},
author = {Hamidian, M. H. and Edkins, S. D. and Joo, Sang Hyun and Kostin, A. and Eisaki, H. and Uchida, S. and Lawler, M. J. and Kim, E.-A. and Mackenzie, A. P. and Fujita, K. and Lee, Jinho and Davis, J. C. Séamus},
journal = {Nature},
volume = {532},
issue = {7599},
pages = {343--347},
numpages = {},
year = {2016},
month = {April},
doi = {10.1038/nature17411},
url = {https://doi.org/10.1038/nature17411}
}

@article{JCDavis19,
author = {S. D. Edkins  and A. Kostin  and K. Fujita  and A. P. Mackenzie  and H. Eisaki  and S. Uchida  and Subir Sachdev  and Michael J. Lawler  and E.-A. Kim  and J. C. Séamus Davis  and M. H. Hamidian },
title = {Magnetic field-induced pair density wave state in the cuprate vortex halo},
journal = {Science},
volume = {364},
number = {6444},
pages = {976--980},
year = {2019},
doi = {10.1126/science.aat1773},
url = {https://www.science.org/doi/abs/10.1126/science.aat1773},
eprint = {https://www.science.org/doi/pdf/10.1126/science.aat1773}
}

@article{bilayerTMD,
title = {Unconventional Superconductivity in Bilayer Transition Metal Dichalcogenides},
author = {Liu, Chao-Xing},
journal = {Phys. Rev. Lett.},
volume = {118},
issue = {8},
pages = {087001},
numpages = {5},
year = {2017},
month = {Feb},
publisher = {American Physical Society},
doi = {10.1103/PhysRevLett.118.087001},
url = {https://link.aps.org/doi/10.1103/PhysRevLett.118.087001}
}

@article{Lu,
author = {Lu, J. M. and Zheliuk, O. and others},
title = {Evidence for two-dimensional Ising superconductivity in gated MoS2},
volume = {350},
number = {6266},
pages = {1353--1357},
year = {2015},
doi = {10.1126/science.aab2277},
publisher = {American Association for the Advancement of Science},
issn = {0036-8075},
url = {https://science.sciencemag.org/content/350/6266/1353},
journal = {Science}
}

@article{Yu,
title = {Superconductivity protected by spin–valley locking in ion-gated MoS2},
author = {Saito, Yu and Nakamura, Yasuharu and others},
journal = {Nat. Phys.},
volume = {12},
issue = {2},
pages = {144--149},
year = {2016},
month = {Feb},
doi = {10.1038/nphys3580},
url = {https://doi.org/10.1038/nphys3580}
}

@article{Xi,
title = {Ising pairing in superconducting NbSe2 atomic layers},
author = {X. Xi and Z. Wang and others},
journal = {Nat. Phys.},
volume = {12},
issue = {2},
pages = {139--143},
year = {2016},
month = {Feb},
doi = {10.1038/nphys3538},
url = {https://doi.org/10.1038/nphys3538}
}

@article{YangKun,
author = {Yang, Kun and Sondhi, S. L.},
title = {Zeeman and orbital effects of an in-plane magnetic field in cuprate superconductors},
journal = {J. Appl. Phys.},
volume = {87},
number = {9},
pages = {5549--5551},
year = {2000},
doi = {10.1063/1.373400},
url = {https://doi.org/10.1063/1.373400}
}

@inproceedings{LawrenceDoniach,
author = {W. E. Lawrence and S. Doniach},
booktitle = {Proc. 12th Int. Conf. Low Temp. Phys.},
pages = {361},
editor = {E. Kanda},
publisher = {Academic},
address = {Kyoto},
year = {1971}
}

@article{Zheng21,
author = {Zhen Zhu  and Michał Papaj  and Xiao-Ang Nie  and Hao-Ke Xu  and Yi-Sheng Gu  and Xu Yang  and Dandan Guan  and Shiyong Wang  and Yaoyi Li  and Canhua Liu  and Jianlin Luo  and Zhu-An Xu  and Hao Zheng  and Liang Fu  and Jin-Feng Jia },
title = {Discovery of segmented Fermi surface induced by Cooper pair momentum},
journal = {Science},
volume = {374},
number = {6573},
pages = {1381--1385},
year = {2021},
doi = {10.1126/science.abf1077}
}

@article{PDW1144,
title = {Smectic pair-density-wave order in EuRbFe4As4},
author = {Zhao, He and Blackwell, Raymond and Thinel, Morgan and Handa, Taketo and Ishida, Shigeyuki and Zhu, Xiaoyang and Iyo, Akira and Eisaki, Hiroshi and Pasupathy, Abhay N and Fujita, Kazuhiro},
journal = {Nature},
volume = {618},
number = {7967},
pages = {940--945},
year = {2023},
publisher = {Nature Publishing Group UK London}
,
doi = {10.1038/s41586-023-06160-5}
,
url = {https://doi.org/10.1038/s41586-023-06160-5}
}

@article{Cao2016,
title = {Superconductivity and ferromagnetism in hole-doped ${\mathrm{RbEuFe}}_{4}{\mathrm{As}}_{4}$},
author = {Liu, Yi and Liu, Ya-Bin and Tang, Zhang-Tu and Jiang, Hao and Wang, Zhi-Cheng and Ablimit, Abduweli and Jiao, Wen-He and Tao, Qian and Feng, Chun-Mu and Xu, Zhu-An and Cao, Guang-Han},
journal = {Phys. Rev. B},
volume = {93},
issue = {21},
pages = {214503},
numpages = {9},
year = {2016},
month = {Jun},
publisher = {American Physical Society},
doi = {10.1103/PhysRevB.93.214503},
url = {https://link.aps.org/doi/10.1103/PhysRevB.93.214503}
}

@article{Kawashima2016,
author = {Kawashima ,Kenji and Kinjo ,Tatsuya and Nishio ,Taichiro and Ishida ,Shigeyuki and Fujihisa ,Hiroshi and Gotoh ,Yoshito and Kihou ,Kunihiro and Eisaki ,Hiroshi and Yoshida ,Yoshiyuki and Iyo ,Akira},
title = {Superconductivity in Fe-Based Compound EuAFe4As4 (A = Rb and Cs)},
journal = {J. Phys. Soc. Jpn.},
volume = {85},
number = {6},
pages = {064710},
year = {2016},
doi = {10.7566/JPSJ.85.064710}
}

@article{Iida2019,
title = {Coexisting spin resonance and long-range magnetic order of Eu in ${\mathrm{EuRbFe}}_{4}{\mathrm{As}}_{4}$},
author = {Iida, K. and Nagai, Y. and Ishida, S. and Ishikado, M. and Murai, N. and Christianson, A. D. and Yoshida, H. and Inamura, Y. and Nakamura, H. and Nakao, A. and Munakata, K. and Kagerbauer, D. and Eisterer, M. and Kawashima, K. and Yoshida, Y. and Eisaki, H. and Iyo, A.},
journal = {Phys. Rev. B},
volume = {100},
issue = {1},
pages = {014506},
numpages = {8},
year = {2019},
month = {Jul},
publisher = {American Physical Society},
doi = {10.1103/PhysRevB.100.014506},
url = {https://link.aps.org/doi/10.1103/PhysRevB.100.014506}
}

@article{liu2021iron,
title = {Iron-based magnetic superconductors AEuFe4As4 (A= Rb, Cs): natural superconductor--ferromagnet hybrids},
author = {Liu, Ya-Bin and Liu, Yi and Cao, Guang-Han},
journal = {J. Phys.: Condens. Matter},
volume = {34},
number = {9},
pages = {093001},
year = {2021},
publisher = {IOP Publishing}
,
doi = {10.1088/1361-648X/ac3cf2}
,
url = {https://dx.doi.org/10.1088/1361-648X/ac3cf2}
}

@article{Qiu2022,
title = {Inhomogeneous superconducting states in two weakly linked superconducting ultrathin films},
author = {Qiu, Gao-Wei and Zhou, Yi},
journal = {Phys. Rev. B},
volume = {105},
issue = {10},
pages = {L100506},
numpages = {6},
year = {2022},
month = {Mar},
publisher = {American Physical Society},
doi = {10.1103/PhysRevB.105.L100506},
url = {https://link.aps.org/doi/10.1103/PhysRevB.105.L100506}
}

@article{BilbaoII,
author = "Aroyo, Mois I. and Kirov, Asen and Capillas, Cesar and Perez-Mato, J. M. and Wondratschek, Hans",
title = "{Bilbao Crystallographic Server. II. Representations of crystallographic point groups and space groups}",
journal = {Acta Crystallogr. A},
year = "2006",
volume = "62",
number = "2",
pages = "115--128",
month = "Mar",
doi = {10.1107/S0108767305040286},
url = {https://doi.org/10.1107/S0108767305040286}
}

@article{Bilbao,
title = {Crystallography online: Bilbao Crystallographic Server},
author = {Aroyo, M. I. and Perez-Mato, J. M. and Orobengoa, D. and Tasci, E. and de la Flor, G. and Kirov, A.},
pages = {183--197},
volume = {43},
number = {2},
journal = {Bulg. Chem. Commun.},
doi = {},
year = {2011}
}

@article{BilbaoMag,
author = "Perez-Mato, J.M. and Gallego, S.V. and Tasci, E.S. and Elcoro, L. and de la Flor, G. and Aroyo, M.I.",
title = "Symmetry-Based Computational Tools for Magnetic Crystallography",
journal = {Annu. Rev. Mater. Res.},
year = "2015",
volume = "45",
number = "Volume 45, 2015",
pages = "217-248",
doi = "https://doi.org/10.1146/annurev-matsci-070214-021008",
url = "https://www.annualreviews.org/content/journals/10.1146/annurev-matsci-070214-021008",
publisher = "Annual Reviews",
issn = "1545-4118",
type = "Journal Article"
}

@book{Tinkham2003Group,
author = {Tinkham, Michael},
day = 17,
isbn = {0486432475},
month = dec,
publisher = {Dover Publications},
title = {Group Theory and Quantum Mechanics (Dover Books on Chemistry)},
url = {http://www.worldcat.org/isbn/0486432475},
year = 2003
}

@article{smidman2017rashba,
title = {Superconductivity and spin--orbit coupling in non-centrosymmetric materials: a review},
author = {Smidman, M. and Salamon, M. B. and Yuan, H. Q. and Agterberg, D. F.},
journal = {Rep. Prog. Phys.},
volume = {80},
number = {3},
pages = {036501},
year = {2017},
publisher = {IOP Publishing},
doi = {10.1088/1361-6633/80/3/036501}
}

@article{kaur2005helical,
title = {Helical spin-triplet superconductivity},
author = {Kaur, R. P. and Agterberg, D. F. and Sigrist, M.},
journal = {Phys. Rev. Lett.},
volume = {94},
number = {13},
pages = {137002},
year = {2005},
publisher = {APS},
doi = {10.1103/PhysRevLett.94.137002}
}

@incollection{agterberg2012,
title = {Magnetism and superconductivity in non-centrosymmetric superconductors},
author = {Agterberg, D. F.},
booktitle = {Non-centrosymmetric Superconductors: Introduction and Overview},
editor = {Bauer, Ernst and Sigrist, Manfred},
series = {Lecture Notes in Physics},
volume = {847},
pages = {155--170},
year = {2012},
publisher = {Springer},
doi = {10.1007/978-3-642-24624-1_5}
}

@article{Kim2021,
title = {Electronic structure and coexistence of superconductivity with magnetism in {RbEuFe$_4$As$_4$}},
author = {Kim, T. and Pervakov, K. S. and Evtushinsky, D. V. and Jung, S.-W. and Poelchen, G. and Kummer, K. and Vlasenko, V. A. and Sadakov, A. V. and Usol'tsev, A. S. and Pudalov, V. M. and Roditchev, D. and Stolyarov, V. S. and Vyalikh, D. V. and Borisov, V. and Valent\'{i}, R. and Ernst, A. and Eremeev, S. V. and Chulkov, E. V.},
journal = {Phys. Rev. B},
volume = {103},
pages = {174517},
year = {2021},
doi = {10.1103/PhysRevB.103.174517}
}

@article{Hemmida2021,
title = {Topological magnetic order and superconductivity in {EuRbFe$_4$As$_4$}},
author = {Hemmida, M. and Winterhalter-Stocker, N. and Ehlers, D. and Krug von Nidda, H.-A. and Yao, M. and Bannies, J. and Rienks, E. D. L. and Kurleto, R. and Felser, C. and B\"{u}chner, B. and Fink, J. and Gorol, S. and F\"{o}rster, T. and Arsenijevic, S. and Fritsch, V. and Gegenwart, P.},
journal = {Phys. Rev. B},
volume = {103},
pages = {195112},
year = {2021},
doi = {10.1103/PhysRevB.103.195112}
}

@article{cho2012superconductivity,
title = {Superconductivity of doped Weyl semimetals: Finite-momentum pairing and electronic analog of the ${}^{3}$He-$A$ phase},
author = {Cho, Gil Young and Bardarson, Jens H. and Lu, Yuan-Ming and Moore, Joel E.},
journal = {Phys. Rev. B},
volume = {86},
issue = {21},
pages = {214514},
numpages = {8},
year = {2012},
month = {Dec},
publisher = {American Physical Society},
doi = {10.1103/PhysRevB.86.214514},
url = {https://link.aps.org/doi/10.1103/PhysRevB.86.214514}
}

	\clearpage
	\onecolumngrid
	\begin{center}
		\textbf{\large Supplemental Material for ``A helical Rashba--exchange gauge field drives a uniaxial pair density wave in EuRbFe$_4$As$_4$''}\\[.2cm]
		Pengfei Li$^{1}$ and Yi Zhou$^{2}$\\[.1cm]
		{\itshape \small $^1$Department of Physics, The Hong Kong University of Science and Technology, Clear Water Bay, Kowloon, Hong Kong, China\\
			$^2$Institute of Physics, Chinese Academy of Sciences, Beijing 100190, China}
	\end{center}
	\vspace{0.5cm}
	
	\setcounter{equation}{0}
	\setcounter{figure}{0}
	\setcounter{table}{0}
	\setcounter{page}{1}
	\renewcommand{\thefigure}{S\arabic{figure}}
	\renewcommand{\thetable}{S\Roman{table}}
	
	\appendix
	\setcounter{secnumdepth}{3}
	\numberwithin{equation}{section}
	\renewcommand{\thesubsection}{\Alph{section}.\arabic{subsection}}
	\section{Rashba spin-orbit coupling and the effective $U(1)$ gauge field}
	\label{app:rashba}
	
	\subsection{Broken local inversion symmetry at the FeAs planes}
	EuRbFe$_4$As$_4$ crystallizes in the centrosymmetric space group $P4/mmm$ (No.~123) \cite{Cao2016,Kawashima2016}.  In this group the Fe atoms occupy the $4i$ Wyckoff position,
	\begin{equation}
		(0,\tfrac12,z),\quad (\tfrac12,0,z),\quad (0,\tfrac12,-z),\quad (\tfrac12,0,-z),
	\end{equation}
	with $z\approx0.13$.  Inversion through the origin sends $(0,\tfrac12,z)$ to $(0,\tfrac12,-z)$; the FeAs sheet at height $+z$ is mapped onto a \emph{different} FeAs sheet at $-z$.  A single FeAs plane therefore does \emph{not} coincide with a crystallographic inversion centre.  Global inversion is preserved, but it interchanges two FeAs layers that have opposite chemical environments:
	\begin{center}
		\begin{tabular}{c|cc}
			\hline\hline
			FeAs layer & layer below & layer above \\
			\hline
			$+z$ & Eu$^{2+}$ & Rb$^+$ \\
			$-z$ & Rb$^+$ & Eu$^{2+}$ \\
			\hline\hline
		\end{tabular}
	\end{center}
	Because the neighbouring interstitial sheets are chemically distinct (different charge, different ionic radius), the electrostatic potential $V(z)$ across a given FeAs plane is asymmetric, $\partial_z V\neq0$, and a finite Rashba spin-orbit coupling (SOC) is allowed locally \cite{smidman2017rashba}.
	
	A subtlety merits emphasis: since the global crystal is centrosymmetric, the electric-field gradient reverses sign between the two FeAs layers in the chemical unit cell.  Consequently the bare Rashba parameter alternates,
	\begin{equation}
		\alpha_R^{(+z)}=-\alpha_R^{(-z)}\equiv\alpha_R,
	\end{equation}
	so that the full lattice Hamiltonian remains inversion-even.  In the four-layer magnetic unit cell this sign alternation must be combined with the $90^\circ$ rotation of the Eu exchange field.  We use $\mathbf H_l$ throughout for the exchange field acting on conduction electrons, with magnitude $H_{\rm ex}$ measured as an energy.  The bare layer-resolved sign $\alpha_l$ should not be identified one-to-one with the effective coefficient $\alpha_R$ used in the GL functional: after projection onto the superconducting $d_{xz}/d_{yz}$ doublet and after fixing a layer/orbital basis convention, a layer-dependent sign can be absorbed into the definition of the low-energy order parameter and the interlayer Josephson amplitudes.  Thus $\alpha_R$ below denotes the effective signed Rashba-exchange coefficient whose sign fixes the handedness of Eq.~\eqref{eq:Al}, while its magnitude fixes $k_0$.  Choosing $\mathbf H_l=H_{\rm ex}(\cos l\pi/2,\sin l\pi/2)$ gives the GL convention used in this work:
	\begin{center}
		\renewcommand\arraystretch{1.25}
		\begin{tabular}{c|c|c|c}
			\hline\hline
				$l$ & local $\alpha_l/\alpha_R$ & $\mathbf H_l/H_{\rm ex}$ & $\mathbf A_l/(H_{\rm eff}a)=\mathbf q_l/k_0$ \\
				\hline
				$0$ & $+$ & $\hat{\mathbf x}$ & $-\hat{\mathbf y}$ \\
				$1$ & $-$ & $\hat{\mathbf y}$ & $\hat{\mathbf x}$ \\
				$2$ & $+$ & $-\hat{\mathbf x}$ & $\hat{\mathbf y}$ \\
				$3$ & $-$ & $-\hat{\mathbf y}$ & $-\hat{\mathbf x}$ \\
			\hline\hline
		\end{tabular}
	\end{center}
	Thus the effective gauge field and the GL momentum shift used throughout the paper follow the single period-four sequence $(-\hat{\mathbf y},\hat{\mathbf x},\hat{\mathbf y},-\hat{\mathbf x})$, exactly matching main-text Fig.~\ref{fig1} and Eq.~\eqref{eq:Al} for $H_{\rm eff}>0$.  The microscopic Rashba alternation fixes the possible handedness and magnitude of the Rashba-exchange coupling, but it does not reduce the magnetic unit cell or invalidate the $P_c4_122$ symmetry analysis.  The GL theory therefore takes Eq.~\eqref{eq:Al}, or equivalently the definition $\mathbf q_l=k_0\hat{\mathbf n}_l$ in the main text, as the definition of the effective helical gauge field after this projection and convention choice.
	
	\subsection{Microscopic estimate of the momentum shift}
	In the presence of Rashba SOC and an in-plane exchange field $\mathbf H_l$, the single-particle Hamiltonian for the $l$-th FeAs layer contains
	\begin{equation}
		\mathcal H_{\rm R}=\alpha_R\,(\boldsymbol\sigma\times\mathbf k)\cdot\hat{\mathbf z}
		\quad\text{and}\quad
		\mathcal H_{\rm ex}=-\mathbf H_l\cdot\boldsymbol\sigma .
	\end{equation}
	Treating $\mathcal H_{\rm ex}$ as a perturbation on the Rashba-split bands shifts the two Fermi circles by opposite momenta.  In the effective GL convention fixed by Eq.~\eqref{eq:Al}, the resulting centre-of-mass momentum of a Cooper pair is \cite{kaur2005helical,agterberg2012}
	\begin{equation}
		\label{eq:ql_micro}
		\mathbf q_l\approx-\frac{2\alpha_R}{\hbar^2v_F^2}\,(\hat{\mathbf z}\times\mathbf H_l) .
	\end{equation}
	The magnitude can be expressed in the dimensionless form
	\begin{equation}
		\frac{q_l}{k_F}\approx\frac{(|\alpha_R| k_F)\,H_{\rm ex}}{2E_F^2},
	\end{equation}
	where $E_F$ is the Fermi energy and $\alpha_R k_F$ the Rashba spin-splitting energy.  This expression is perturbative in the exchange field and is controlled when $H_{\rm ex}/E_F$ and $q/k_F$ are small; below we use it only as an order-of-magnitude estimate of the natural modulation scale.  The exchange field $H_{\rm ex}$ on the conduction electrons is set by the Eu--Fe exchange coupling $J_{\rm ex}$ via $H_{\rm ex}(T) = J_{\rm ex}\langle S_{\rm Eu}(T)\rangle$.  In the zero-temperature saturated limit, $\langle S_{\rm Eu}\rangle=S_{\rm Eu}=7/2$.  Since the Eu$^{2+}$--Eu$^{2+}$ magnetic interaction is predominantly mediated by the RKKY mechanism through the Fe~$3d$ conduction electrons~\cite{liu2021iron}, a mean-field estimate gives $k_B T_m \sim J_{\rm ex}^2 N(0)\, S(S+1)/3$, where $N(0)$ is the bare density of states at the Fermi level per spin.  First-principles calculations give $\gamma_{\rm bare}\approx 21.6\,$mJ\,K$^{-2}$\,mol$^{-1}$~\cite{liu2021iron}, corresponding to $N(0)\approx 4.6\,$states/eV per formula unit per spin.  Numerically,
	\begin{equation}
		J_{\rm ex}\sim\sqrt{\frac{3k_BT_m}{N(0)S(S+1)}}
		\approx\sqrt{\frac{3\times1.29\,{\rm meV}}{4.6\,{\rm eV}^{-1}\times15.75}}
		\approx7\,{\rm meV}.
	\end{equation}
	This yields a zero-temperature estimate $H_{\rm ex} \approx 26\,$meV.  For a shallow parabolic pocket, $\hbar v_F\simeq2E_F/k_F$, which connects this form to the main-text estimate using $v_F$.  This RKKY extraction is approximate: it neglects momentum-dependent susceptibility, anisotropic exchange paths, and the reduction of $\langle S_{\rm Eu}\rangle$ near $T_m$.
	
	First-principles calculations~\cite{liu2021iron} show that the $d_{xz}/d_{yz}$ hole pockets at $\Gamma$, which are most relevant for the PDW instability, are notably shallow: $E_F \sim 10$--$15\,$meV and $k_F \sim 0.10$--$0.12\,\text{\AA}^{-1}$.  Combined with a Rashba spin splitting $\alpha_R k_F \sim 10$--$20\,$meV, the perturbative estimate gives $q/k_F \sim 1$--$2$.  This value is at the boundary of the weak-exchange regime, so the numerical wavelength should not be over-interpreted.  It nevertheless indicates that the Rashba-exchange mechanism naturally produces a modulation length of only a few nanometers,
	\begin{equation}
		\lambda=\frac{2\pi}{q}\sim 3\;\text{nm}\approx 8\;a_{\rm Fe},
	\end{equation}
	comparable to the STM observation of $\approx8$ unit-cell modulation \cite{PDW1144}.  Quantitative refinement of this number requires a non-perturbative calculation using the actual multiband Fermi surfaces.
	
	\subsection{Effective $U(1)$ gauge field}
	In the Ginzburg-Landau framework a finite-momentum shift acts as a covariant derivative,
	\begin{equation}
		\mathbf D_l=\boldsymbol\nabla-i\mathbf q_l ,
	\end{equation}
	with $\mathbf q_l=k_0\hat{\mathbf n}_l$ and $k_0=2|\alpha_R|H_{\rm ex}/(\hbar^2v_F^2)>0$ in the effective GL convention. This is mathematically identical to the minimal coupling of a $U(1)$ gauge potential
	\begin{equation}
		\mathbf A_l^{\rm eff}=\frac{\hbar c}{2e}\,\mathbf q_l
		\equiv H_{\rm eff}\,a\Bigl(\sin\frac{l\pi}{2}\,\hat{\mathbf x}-\cos\frac{l\pi}{2}\,\hat{\mathbf y}\Bigr) .
	\end{equation}
	The formal effective-field scale defined in the main text is therefore
	\begin{equation}
		H_{\rm eff}=\frac{\hbar c}{2ea}k_0
		=\frac{|\alpha_R|\,c\,H_{\rm ex}}{e\,\hbar\,v_F^2\,a}>0 ,
	\end{equation}
	where the positive scale is paired with the convention $\mathbf q_l\propto-\hat{\mathbf z}\times\mathbf H_l$. Equivalently, $\bigl(\sin\frac{l\pi}{2},-\cos\frac{l\pi}{2}\bigr)=-\hat{\mathbf z}\times\bigl(\cos\frac{l\pi}{2},\sin\frac{l\pi}{2}\bigr)$, exactly as used in Eq.~(3).  The variables $\mathbf A_l^{\rm eff}$ and $H_{\rm eff}$ are formal gauge-field variables; the physical input is the momentum shift $\mathbf q_l$, or equivalently $k_0$. Because the Eu$^{2+}$ spins rotate by $90^\circ$ between adjacent magnetic layers, the induced gauge field likewise rotates by $90^\circ$ along the $c$-axis, strictly satisfying the periodic condition $\mathbf A_{l+4}^{\rm eff}=\mathbf A_l^{\rm eff}$.
	
	The magnitude of $H_{\rm eff}$ can be estimated directly.  Using $\alpha_R\sim 0.1\,$eV\,\AA, $H_{\rm ex}\approx 26\,$meV, $v_F\sim 3$--$4\times10^4\,$m/s (corresponding to $E_F\sim 10$--$15\,$meV for the shallow $d_{xz}/d_{yz}$ pockets~\cite{liu2021iron}) and $a\sim 1.3\,$nm, one finds $H_{\rm eff}\sim 500$--$2000\,$T.  This large number should be understood only as a formal conversion of the momentum shift into magnetic-field units through $\mathbf A_l^{\rm eff}=(\hbar c/2e)\mathbf q_l$.  It is neither a real magnetic induction nor an orbital depairing field.  The physically meaningful comparison is instead between length scales: the Rashba-exchange momentum shift gives a nanometer-scale modulation, whereas the bare dipolar orbital effect of the ordered Eu moments would give a micron-scale modulation.

	\section{Symmetry of the SC order parameter}
	The structural symmetry of EuRbFe$_4$As$_4$ corresponds to the space group $P4/mmm$ (No.~123) with Patterson symmetry $P4/mmm$. However, when the $90^\circ$ helical magnetic order forms below $T_m$, the system possesses an enlarged $4a$ magnetic unit cell and is governed by the magnetic space group $P_{c}4_{1}22$ (No.~91.108), which is a type IV Shubnikov group (black-white groups with black-white Bravais lattices)~\cite{Bilbao,BilbaoMag}. This magnetic space group has the space group $P4_{1}22$ (No.~91) as a subgroup, which shares the Patterson symmetry $P4/mmm$ with space group No.~123. 
	
	The space group $P4_{1}22$ includes eight symmetry operations. Using Seitz notation~\cite{Tinkham2003Group}, $\{\alpha|\mathbf{t}\}\mathbf{r}=\alpha\mathbf{r}+\mathbf{t}$ (where $\alpha$ denotes a point group operation and $\mathbf{t}$ is a translational vector), these operations are identified as $\{1|0,0,0\}$, $\{2_{001}|0,0,1/2\}$, $\{4^+_{001}|0,0,1/4\}$, $\{4^-_{001}|0,0,3/4\}$, $\{2_{010}|0,0,0\}$, $\{2_{100}|0,0,1/2\}$, $\{2_{110}|0,0,3/4\}$, and $\{2_{1\bar{1}0}|0,0,1/4\}$. These can be derived from the two fundamental generators $\{4^+_{001}|0,0,1/4\}$ and $\{2_{010}|0,0,0\}$. 
	The magnetic space group $P_{c}4_{1}22$ is connected to the space group $P4_{1}22$ by the relation $P_{c}4_{1}22=P4_{1}22+\mathcal{T}\{0|0,0,1/2\}P4_{1}22$, where $\mathcal{T}$ is the time-reversal operation. 
	
	We first examine the irreducible representations of the space group $P4_{1}22$, which strictly constrain the SC order parameter $\psi_{l}(\mathbf{r})$. Defining the enlarged Bloch unit cell dimensions as $a_0\times a_0\times 4a$, the Bloch wave function for the SC ground state emerges at the $\Gamma$ point [wave vector $\mathbf{k}=(0,0,0)$]. According to the irreducible representations (irreps) provided by the Bilbao Crystallographic Server~\cite{Bilbao,BilbaoII}, we obtain the constraints for $\psi_{l}(\mathbf{r})$ in each of the five irreducible representations, $\Gamma_{1,2,3,4,5}$. These are determined by the two symmetry generators $\{4^+_{001}|0,0,1/4\}$ and $\{2_{010}|0,0,0\}$.
	
	Next, we incorporate the symmetry operation $\mathcal{T}\{0|0,0,1/2\}$. Because the SC order parameter characterizes a condensate of electron pairs, it must satisfy $\mathcal{T}^2=1$ at the $\Gamma$ point. According to Wigner's theorem~\cite{Tinkham2003Group}, the relationship between an irreducible representation $\Gamma(R)$ and its complex conjugate $\Gamma(R)^{\ast}$ can either be (a) $\Gamma(R)=\Gamma(R)^{\ast}$ (a real representation), or (b) $\Gamma(R)$ is complex and not equivalent to $\Gamma(R)^{\ast}$. In the magnetic space group $P_{c}4_{1}22$, which is associated with the crystal point group $D_4$, only real irreducible representations are allowed at the $\Gamma$ point.
	
	\subsection{Symmetry Constraints and Table of Irreducible Representations}
	The symmetry operations and the corresponding irreducible representations of the magnetic space group $P_{c}4_{1}22$ impose strict constraints on the layer-dependent SC order parameter. Table~\ref{tab:SCorderSM} summarizes these constraints for both the 1D representations ($\Gamma_1$ to $\Gamma_4$) and the 2D representation ($\Gamma_5$) discussed in the main text.
	
	\begin{table}[h]
		\caption{Five real irreducible representations for the magnetic space group $P_{c}4_{1}22$ at the $\Gamma$ point~\cite{Bilbao,BilbaoII}, imposing symmetry constraints on the SC order parameter. Here $\sigma_x, \sigma_y, \sigma_z$ are Pauli matrices. The periodic condition $\psi_{l+4}(x,y)=\psi_{l}(x,y)$ is applied. In addition, the constraint $\psi_{l+2}(x,y)^{\ast}=\psi_{l}(x,y)$ is imposed by $\mathcal{T}\{0|0,0,1/2\}$.}
		\renewcommand\arraystretch{1.5}
		\setlength{\tabcolsep}{0.9ex}
		\begin{tabular}{c|ccccc}
			\hline\hline
			irreps. & $\Gamma_1$ & $\Gamma_2$ & $\Gamma_3$ & $\Gamma_4$ & $\Gamma_5$ \\ 
			\hline
			\multirow{3}{*}{$\{4^+_{001}|0,0,\frac{1}{4}\}$}  & $1$ & $-1$ & $1$ & $-1$ & $-i\sigma_y$  \\
			\cline{2-6}
			& \multicolumn{4}{c}{$\psi_{l+1}(-y,x)=\eta\,\psi_{l}(x,y)$} & $\begin{array}{l}
				\psi_{l+1,1}(-y,x)=-\psi_{l,2}(x,y)\\
				\psi_{l+1,2}(-y,x)=\psi_{l,1}(x,y)
			\end{array}$ \\
			\hline
			\multirow{3}{*}{$\{2_{010}|0,0,0\}$}  & $1$ & $1$ & $-1$ & $-1$ & $-\sigma_z$\\
			\cline{2-6}
			& \multicolumn{4}{c}{$\psi_{-l}(-x,y)=\zeta\,\psi_{l}(x,y)$} & $\begin{array}{l}
				\psi_{-l,1}(-x,y)=-\psi_{l,1}(x,y)\\
				\psi_{-l,2}(-x,y)=\psi_{l,2}(x,y)
			\end{array}$\\
			\hline\hline
		\end{tabular}
		\label{tab:SCorderSM}
	\end{table}
	
	For the Bloch SC states, it is straightforward to verify that the phase factor associated with the effective gauge field, $e^{i\frac{2\pi}{\Phi_0} \mathbf{A}_{l}\cdot\mathbf{r}}=e^{ik_0\left(\sin\frac{l\pi}{2}x-\cos\frac{l\pi}{2}y\right)}$, is invariant under all symmetry transformations of $P_{c}4_{1}22$. Consequently, the periodic amplitude function $\tilde{\psi}_l(x,y)$ must transform as:
	\begin{equation}\label{eq:tranpsi}
		\begin{split}
			\{4^+_{001}|0,0,1/4\}&:\tilde{\psi}_{l}(x,y) =\eta\,\tilde{\psi}_{l+1}(-y,x),\\
			\{2_{010}|0,0,0\}&:\tilde{\psi}_{l}(x,y)=\zeta\,\tilde{\psi}_{-l}(-x,y), \\
			\mathcal{T}\{0|0,0,1/2\}&:\tilde{\psi}_{l}(x,y)=\tilde{\psi}_{l+2}(x,y)^{\ast}.
		\end{split}
	\end{equation}
	For the four 1D representations $\Gamma_{1,2,3,4}$, expanding $\tilde{\psi}_l$ in Fourier space allows the last equation in Eq.~\eqref{eq:tranpsi} to yield:
	\begin{subequations}\label{eq:aluv}
		\begin{equation}
			a_{l+2,-\mu,-\nu}^{\ast}=a_{l,\mu,\nu},    
		\end{equation}
		while the other two equations impose:
		\begin{eqnarray}
			a_{l+1,-\nu,\mu} & = & \eta{}a_{l,\mu,\nu}, \\
			a_{-l,-\mu,\nu} & = & \zeta{}a_{l,\mu,\nu},   
		\end{eqnarray}
	\end{subequations}
	where $\eta=\pm{}1$ and $\zeta=\pm{}1$ denote the characters of $\{4^+_{001}|0,0,1/4\}$ and $\{2_{010}|0,0,0\}$, respectively.

	\section{In-plane gradient free energy $f_{\mathrm{grad}}$ for the 2D irrep.\ $\Gamma_5$}
	The Ginzburg-Landau free energy density must be invariant under the point-group operations of the crystal (here $D_4$) and under local $U(1)$ gauge transformations. We consider a two-component order parameter $\vec{\psi}=(\psi_1,\psi_2)$ transforming as the 2D irreducible representation $\Gamma_5$ of $D_4$. 
	
	Gradient terms are built from the covariant derivatives:
	\begin{equation}
		D_j = \partial_j - i \frac{2\pi}{\Phi_0} A_j, \qquad j=x,y.
	\end{equation}
	In $D_4$, both $(D_x,D_y)$ and $(\psi_1,\psi_2)$ transform as $\Gamma_5$. Therefore, quadratic gradient terms that are bilinear in $\psi$ can be organized by forming $D_4$-invariants from the product $\Gamma_5\otimes\Gamma_5=\Gamma_1\oplus\Gamma_2\oplus\Gamma_3\oplus\Gamma_4$, yielding four independent invariants.
	
	\subsection{Most general $D_4$-invariant in-plane gradient energy}
	The most general in-plane contribution quadratic in both gradients and $\psi$ can be written as:
	\begin{subequations}\label{eq:fgrad_general}
		\begin{align}
			f_{\mathrm{grad}}
			&=K_1\Bigl(|D_x\psi_1|^2+|D_y\psi_2|^2\Bigr) \label{eq:K1}\\
			&\quad+K_2\Bigl(|D_y\psi_1|^2+|D_x\psi_2|^2\Bigr) \label{eq:K2}\\
			&\quad+K_3\Bigl[(D_x\psi_1)^{*}(D_y\psi_2)+(D_y\psi_1)^{*}(D_x\psi_2)+\text{c.c.}\Bigr] \label{eq:K3}\\
			&\quad+K_4\Bigl[(D_x\psi_1)^{*}(D_y\psi_2)-(D_y\psi_1)^{*}(D_x\psi_2)+\text{c.c.}\Bigr]. \label{eq:K4}
		\end{align}
	\end{subequations}
	Here ``c.c.'' denotes the complex conjugate. The $K_3$ and $K_4$ terms represent two $D_4$-symmetry-allowed mixed-gradient invariants. 
	
	\subsection{Useful soft-mode limit}
	The main text uses the analytically transparent $D_4$-symmetric soft-mode choice $K_1=K_2=K_3\equiv K$ and $K_4=0$.  This limit is not required by symmetry; the generic coefficients in Eq.~\eqref{eq:fgrad_general} remain independent phenomenological parameters subject only to stability.  Its value is that it exposes the uniaxial instability in closed form, while generic $D_4$ stiffnesses lift the exact zero modes and shift the phase boundaries without changing the symmetry-allowed nature of the finite-momentum state.
	
	\textbf{Stability analysis.}
	Introducing the shorthand $a=D_x\psi_1$, $b=D_y\psi_1$, $c=D_x\psi_2$, $d=D_y\psi_2$, the gradient energy~\eqref{eq:fgrad_general} can be written as the Hermitian form $f_{\mathrm{grad}}=\mathbf{v}^\dagger M\,\mathbf{v}$ with $\mathbf{v}=(a,b,c,d)^T$ and
	\begin{equation}\label{eq:Mgrad}
		M=\begin{pmatrix}
			K_1 & 0 & 0 & K_3+K_4\\
			0 & K_2 & K_3-K_4 & 0\\
			0 & K_3-K_4 & K_2 & 0\\
			K_3+K_4 & 0 & 0 & K_1
		\end{pmatrix}.
	\end{equation}
	This matrix block-diagonalizes into two $2\times2$ sectors:
	\begin{itemize}
		\item $(a,d)$ block with eigenvalues $K_1\pm(K_3+K_4)$,
		\item $(b,c)$ block with eigenvalues $K_2\pm(K_3-K_4)$.
	\end{itemize}
	Requiring all four eigenvalues to be strictly positive yields the necessary and sufficient stability conditions:
	\begin{equation}\label{eq:stability_correct}
		K_{1,2}>0,\qquad |K_3+K_4|<K_1,\qquad |K_3-K_4|<K_2.
	\end{equation}
	
	\textbf{Remark on the main-text choice.}
	The simplified choice $K_1=K_2=K_3\equiv K$, $K_4=0$ adopted in the main text saturates the stability bound~\eqref{eq:stability_correct}: the $(a,d)$-block eigenvalue $K_1-(K_3+K_4)=0$, and likewise for the $(b,c)$ block, so $f_{\mathrm{grad}}$ is positive-\emph{semi}definite with two zero modes.  These zero modes are \emph{not} an instability; they reflect an additional symmetry of this parameter choice, as we now show.
	
	Substituting $K_1=K_2=K_3=K$, $K_4=0$ into Eq.~\eqref{eq:fgrad_general} yields the identity
	\begin{equation}\label{eq:fgrad_divform}
		f_{\mathrm{grad}}\big|_{K_1=K_2=K_3=K,\,K_4=0}
		= K\,\bigl|\vec D\cdot\vec\psi\bigr|^2
		+ K\,\bigl|\vec D\cdot(\sigma_x\vec\psi)\bigr|^2,
	\end{equation}
	which is precisely the gradient term appearing in the main-text free energy.  To expose the symmetry, we pass to the rotated basis
	\begin{equation}\label{eq:rotated_basis}
		\phi_{\pm}=\frac{\psi_1\pm\psi_2}{\sqrt{2}},\qquad
		u=\frac{x+y}{\sqrt{2}}\;\bigl(\parallel[110]\bigr),\qquad
		v=\frac{x-y}{\sqrt{2}}\;\bigl(\parallel[1\bar10]\bigr).
	\end{equation}
	A direct substitution gives
	\begin{equation}\label{eq:fgrad_decoupled}
		f_{\mathrm{grad}}=2K\,|D_u\phi_+|^2+2K\,|D_v\phi_-|^2,
	\end{equation}
	where $D_{u,v}$ denote the covariant derivatives along $u$ and $v$.  The gradient energy thus \emph{completely decouples}: $\phi_+$ has stiffness only along the $[110]$ diagonal and is free along $[1\bar10]$, while $\phi_-$ has stiffness only along $[1\bar10]$ and is free along $[110]$.
	
	\textbf{Symmetry origin of the zero modes.}
	The decoupled form~\eqref{eq:fgrad_decoupled} is manifestly invariant under the infinite-dimensional group of shifts
	\begin{equation}\label{eq:shift_symmetry}
		\phi_+(u,v)\to\phi_+(u,v)+\chi(v),\qquad
		\phi_-(u,v)\to\phi_-(u,v)+\xi(u),
	\end{equation}
	for \emph{arbitrary} functions $\chi(v)$ and $\xi(u)$, since $D_u\chi(v)=0$ and $D_v\xi(u)=0$.  In the original basis this reads
	\begin{equation}\label{eq:shift_original}
		\psi_1\to\psi_1+\tfrac{1}{\sqrt{2}}\bigl[\chi((x{-}y)/\sqrt{2})+\xi((x{+}y)/\sqrt{2})\bigr],\qquad
		\psi_2\to\psi_2+\tfrac{1}{\sqrt{2}}\bigl[\chi((x{-}y)/\sqrt{2})-\xi((x{+}y)/\sqrt{2})\bigr].
	\end{equation}
	The two zero eigenvalues of the matrix~\eqref{eq:Mgrad} are the consequence of this accidental shift symmetry of the simplified gradient sector: modulations of $\phi_+$ along $[1\bar10]$ and of $\phi_-$ along $[110]$ cost exactly zero gradient energy.
	
	In momentum space the picture is equally transparent.  For a plane-wave perturbation $\delta\vec\psi\propto(f_1,f_2)\,e^{i\mathbf{q}\cdot\mathbf{r}}$, the two stiffness eigenvalues are $K(q_x+q_y)^2$ and $K(q_x-q_y)^2$: the mode $(1,-1)$ is soft for $\mathbf{q}\parallel[110]$, and the mode $(1,1)$ is soft for $\mathbf{q}\parallel[1\bar10]$.
	
	Because these zero modes are protected by the shift symmetry~\eqref{eq:shift_symmetry}, they do not signal a thermodynamic instability.  The symmetry is not a microscopic space-group symmetry; it is an analytically useful soft-mode limit of the $D_4$ gradient sector.  In the full free energy the quartic and Josephson couplings break this infinite-dimensional symmetry and generically lift the flat directions, selecting a definite ground state.  The choice $K_1=K_2=K_3$, $K_4=0$ is therefore a controlled phenomenological simplification for exposing the uniaxial instability.
	
	If one wishes to make the gradient sector \emph{strictly} positive-definite on its own (e.g.\ for numerical stability of gradient-flow solvers), a convenient one-parameter family that preserves $D_4$ symmetry is
	\begin{equation}\label{eq:K_recommended}
		K_1=K_2=K,\qquad K_3=\eta\,K,\qquad K_4=0,\qquad 0\le\eta<1,
	\end{equation}
	where $\eta\to 1$ recovers the main-text limit.  Any $\eta<1$ introduces a nonzero curl stiffness and lifts the two zero modes while maintaining the $C_4$-symmetric structure $K_1=K_2$.  We have checked representative values of this family in the numerical minimization: reducing $\eta$ shifts the FF--Bloch boundary to larger effective field, as expected from the added stiffness, but the Bloch state that emerges remains uniaxial.  Thus the exact zero modes are not required for the existence of the PDW; they provide the clearest analytic route to the instability.
	
	\section{Minimization of the quartic free energy for the decoupled $\Gamma_5$ spinor}
	\label{app:spinor}
	The ground-state spinor alignment of the decoupled SC solution follows from minimizing the quartic part of the uniform GL free energy,
	\begin{equation}
		\label{eq:quartic_spinor}
		\frac{\beta_1}{2}\bigl(\vec\psi^\dagger\vec\psi\bigr)^2
		+\frac{\beta_2}{2}\bigl(\vec\psi^\dagger\sigma_y\vec\psi\bigr)^2
		+\frac{\beta_3}{2}\bigl(\vec\psi^\dagger\sigma_z\vec\psi\bigr)^2 .
	\end{equation}
	Writing $\vec\psi=\sqrt{\rho}\,(\cos\theta,\sin\theta e^{i\phi})^T$, the normalized quartic energy is
	\begin{equation}
		\label{eq:f_theta_phi}
		f(\theta,\phi)=\beta_1+\beta_2\sin^2(2\theta)\sin^2\phi+\beta_3\cos^2(2\theta).
	\end{equation}
	The global minimum is selected from three candidate spinors,
	\begin{equation}
		\label{eq:spinor_cases}
		\begin{array}{c|c|c}
			\text{state} & (u,v) & f_{\rm min} \\
			\hline
			\text{diagonal nematic} & (1,\pm 1)/\sqrt{2} & \beta_1 \\
			\text{principal nematic} & (1,0)\ \text{or}\ (0,1) & \beta_1+\beta_3 \\
			\text{chiral} & (1,\pm i)/\sqrt{2} & \beta_1+\beta_2
		\end{array}
	\end{equation}
	with thermodynamic stability requiring $\beta_1>0$, $\beta_2> -\beta_1$, and $\beta_3> -\beta_1$.  Comparing the three entries in Eq.~\eqref{eq:spinor_cases} gives the phase boundaries summarized in Table~\ref{tableDSC}.
	
	\begin{table}[tb]
		\caption{Parameters for decoupled SC states in the 2D representation $\Gamma_5$. Thermodynamic stability requires $\beta_1>0$, $\beta_2> -\beta_1$, and $\beta_{3}> -\beta_1$. The spinor $(u_l,v_l)$ is determined up to an overall phase factor. Here, $\bar{\beta}$ is the effective quartic coefficient such that the uniform superfluid density is given by $\rho_s = -\alpha / \bar{\beta}$.}
		\renewcommand\arraystretch{2.0}
		\setlength{\tabcolsep}{1.0ex}
		\begin{tabular}{cccccc}
			\hline\hline
			$\beta_{2}$ & $\beta_{3}$ & Conditions & $(u_{l},v_{l})$ & State & $\bar{\beta}$ \\ \hline
			$+$ & $+$ & & $(1,\,\pm{}1)/\sqrt{2}$ & Diagonal Nematic & $\beta_1$ \\ \hline
			$+$ & $-$ & & $(1,0)$ or $(0,1)$ & Principal Nematic & $\beta_1-|\beta_{3}|$ \\ \hline
			$-$ & $+$ & & $(1,\pm{}i)/\sqrt{2}$ & Chiral & $\beta_1-|\beta_{2}|$ \\ \hline
			$-$ & $-$ & $|\beta_{2}|>|\beta_{3}|$ & $(1,\pm{}i)/\sqrt{2}$ & Chiral & $\beta_{1}-|\beta_{2}|$ \\ \hline
			$-$ & $-$ & $|\beta_{2}|<|\beta_{3}|$ & $(1,0)$ or $(0,1)$ & Principal Nematic & $\beta_{1}-|\beta_{3}|$ \\
			\hline\hline
		\end{tabular}
		\label{tableDSC}
	\end{table}

	\section{Josephson coupling of 2D representation $\Gamma_5$}
	Suppressing the common layer-index shift and coordinate transformation, the internal two-component vector of a basis function $\vec{\psi}_{l}$ in the representation $\Gamma_5$ transforms as follows:
	\begin{equation}\label{eq:gamma5vec}
		\begin{split}
			\{4^+_{001}|0,0,1/4\}&:\vec{\psi}_{l}(x,y) \to (-i\sigma_y)\vec{\psi}_{l}(x,y),\\
			\{2_{010}|0,0,0\}&:\vec{\psi}_{l}(x,y) \to -\sigma_z\vec{\psi}_{l}(x,y), \\
			\mathcal{T}\{0|0,0,1/2\}&:\vec{\psi}_{l}(x,y) \to \vec{\psi}_{l+2}(x,y)^{\ast}.
		\end{split}
	\end{equation}
	Thus, for a general $2\times{}2$ matrix $M = m_0\sigma_0+\vec{m}\cdot\vec{\sigma}$, the inter-layer quadratic coupling term $\vec{\psi}_{l}^{\dagger}M\vec{\psi}_{l+1}$ transforms as:
	\begin{equation}\label{eq:gamma5M}
		\begin{split}
			\{4^+_{001}|0,0,1/4\}&: \vec{\psi}_{l}^{\dagger}M\vec{\psi}_{l+1} \to \vec{\psi}_{l}^{\dagger} (i\sigma_y) M (-i\sigma_y)\vec{\psi}_{l+1},\\
			\{2_{010}|0,0,0\}&: \vec{\psi}_{l}^{\dagger}M\vec{\psi}_{l+1} \to \vec{\psi}_{l}^{\dagger} (-\sigma_{z}) M (-\sigma_z)\vec{\psi}_{l+1}, \\
			\mathcal{T}\{0|0,0,1/2\}&: \vec{\psi}_{l}^{\dagger}M\vec{\psi}_{l+1} \to \vec{\psi}_{l+2}^{T}M^{\ast}\vec{\psi}_{l+3}^{\ast}=\vec{\psi}_{l+3}^{\dagger}M^{\dagger}\vec{\psi}_{l+2}.
		\end{split}
	\end{equation}
	The first line dictates that a $P4_{1}22$-invariant quadratic term $\vec{\psi}_{l}^{\dagger}M\vec{\psi}_{l+1}$ must obey $M=\sigma_0$ or $M=i\sigma_y$, both of which are strictly real matrices as required by the $\mathcal{T}$ operation. 
	
	The choice $M=\sigma_0$ satisfies the second symmetry constraint. We can explicitly verify that $M=i\sigma_y$ also satisfies it through a straightforward shifting of layer indices:
	\begin{equation}
		\begin{aligned}
			& \sum_{l=0}^3 \left( i\vec{\psi}_{l}^{\dagger}\sigma_y\vec{\psi}_{l+1} - i\vec{\psi}_{l+1}^{\dagger}\sigma_y\vec{\psi}_{l} \right) \\
			\to & \sum_{l=0}^3 \left( i\vec{\psi}_{-l}^{\dagger}\sigma_z\sigma_y\sigma_z\vec{\psi}_{-(l+1)} - i\vec{\psi}_{-(l+1)}^{\dagger}\sigma_z\sigma_y\sigma_z\vec{\psi}_{-l} \right) \\
			=& \sum_{l=0}^3 \left( - i\vec{\psi}_{l+1}^{\dagger}\sigma_y\vec{\psi}_{l} + i\vec{\psi}_{l}^{\dagger}\sigma_y\vec{\psi}_{l+1} \right).
		\end{aligned}
	\end{equation}
	The final equality follows from the fact that the order parameter remains invariant under a translation by 4 layers.
	
	\section{GL Differential Equations and Perturbative Onset of the Bloch State}
	By minimizing the GL free energy with respect to $\vec{\psi}_l^\dagger$, we obtain the full GL differential equations for the 2D representation $\Gamma_5$:
	\begin{equation}\label{eq:GLE2_SM}
		\begin{split}
			-\frac{\hbar^2}{2m^*} \left( \mathbf{D}_l^2\mathbb{I} + 2D_{lx}D_{ly}\sigma_x \right)\vec{\psi}_l + \alpha \vec{\psi}_l 
			- g_0 (\vec{\psi}_{l-1} + \vec{\psi}_{l+1}) - i\sigma_y g_2 (\vec{\psi}_{l+1} - \vec{\psi}_{l-1}) \\
			+ \left[ \beta_1 (\vec{\psi}_l^\dagger \vec{\psi}_l) + \beta_2 (\vec{\psi}_l^\dagger \sigma_y \vec{\psi}_l) \sigma_y + \beta_3 (\vec{\psi}_l^\dagger \sigma_z \vec{\psi}_l) \sigma_z \right] \vec{\psi}_l = 0.
		\end{split}
	\end{equation}
	
	To analyze the perturbative onset of the spatially modulated Bloch state, we expand the periodic envelope $\vec{\varphi}_l(\mathbf{r})$ in Fourier harmonics:
	\begin{equation}
		\vec{\varphi}_l(\mathbf{r}) = \sqrt{n}\sum_{\mu,\nu} \begin{pmatrix} a_{l,\mu,\nu} \\ b_{l,\mu,\nu} \end{pmatrix} e^{ik_0(\mu x + \nu y)},
	\end{equation}
	with integer indices $\mu,\nu$. The symmetry constraints relate the Fourier coefficients across layers such that $\vec{\varphi}_{l+1}$ coefficients are determined entirely by applying $-i\sigma_y$ to the rotated coordinates of $\vec{\varphi}_l$.
	
	When small Josephson coupling is introduced, it acts as a perturbation that couples the uniform $(\mu,\nu)=(0,0)$ Fourier component on layer $l$ to first diagonal harmonics on layer $l+1$, with integer shifts set by the difference between the adjacent gauge-field directions. The gradient stiffness for a harmonic $(\mu,\nu)$ is a $2\times2$ matrix in spinor space:
	\begin{equation}
		\mathcal{M}_{(\mu,\nu)}=Kk_0^2\begin{pmatrix}\mu^2+\nu^2 & 2\mu\nu \\ 2\mu\nu & \mu^2+\nu^2\end{pmatrix}.
	\end{equation}
	For the same-sign diagonal harmonics $(\pm1,\pm1)$, one eigenvalue is $4Kk_0^2$ (stiff mode, spinor $(1,1)/\sqrt{2}$) and the other is exactly zero (soft mode, spinor $(1,-1)/\sqrt{2}$). The Josephson source driving the $(-1,-1)$ harmonic on layer $l+1$ is given by:
	\begin{equation}
		\mathbf{S}_{l}=g_0\begin{pmatrix}u_l\\v_l\end{pmatrix}-g_2\,(i\sigma_y)\begin{pmatrix}u_l\\v_l\end{pmatrix}.
	\end{equation}
	Projecting this source vector $\mathbf{S}_l$ onto the zero-energy soft mode dictates the critical threshold for the Bloch instability for each respective decoupled state, as discussed in the main text.
	
	\section{Phase Diagrams}
	This section provides the full parameter space phase diagrams detailing the stability regimes of the decoupled SC, Fulde-Ferrell (FF), and Bloch SC states discussed in the main text. In the numerical plots we use the dimensionless effective-field parameter $h\equiv\hbar^2k_0^2/(2m^*|\alpha|)$ and measure Josephson couplings in units of $|\alpha|$.
	
	The resulting phase boundary between the FF and Bloch SC regimes is approximately described by $h_c \propto \sqrt{g}$. This square-root behavior may have a simple energetic origin: the effective gauge field contributes through the gradient energy and hence scales as $h^2$, whereas the Josephson phase-locking energy is linear in $g$. The transition is therefore expected when $h_c^2$ becomes comparable to $g$.
	
	\begin{figure}[h]
		\begin{center}
			\includegraphics[width=0.5\linewidth]{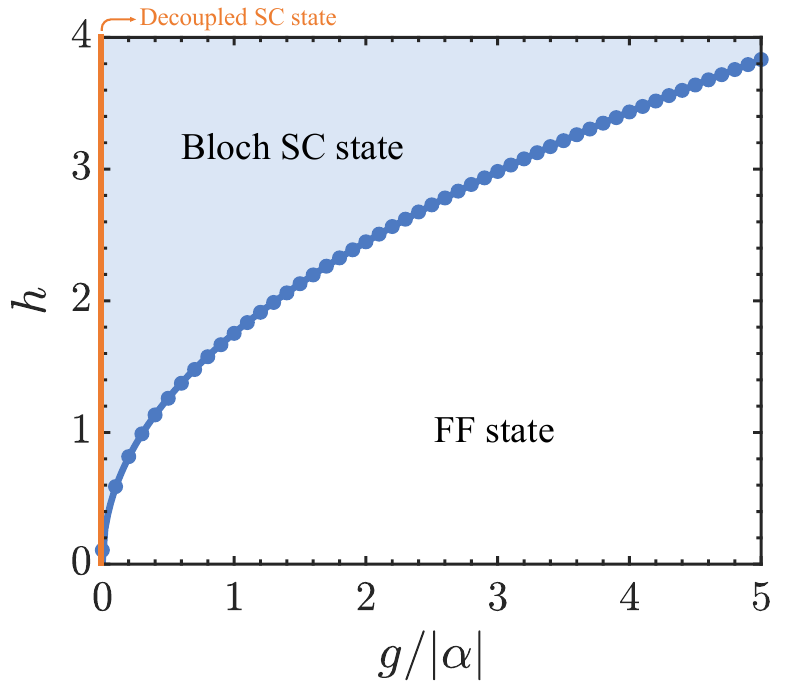}
			\caption{Phase diagram for the 1D representations.}\label{figSM_1Dphase}
		\end{center}
	\end{figure}
	
	\begin{figure}[h]
		\begin{center}
			\subfigure[]{
				\includegraphics[width=0.30\linewidth]{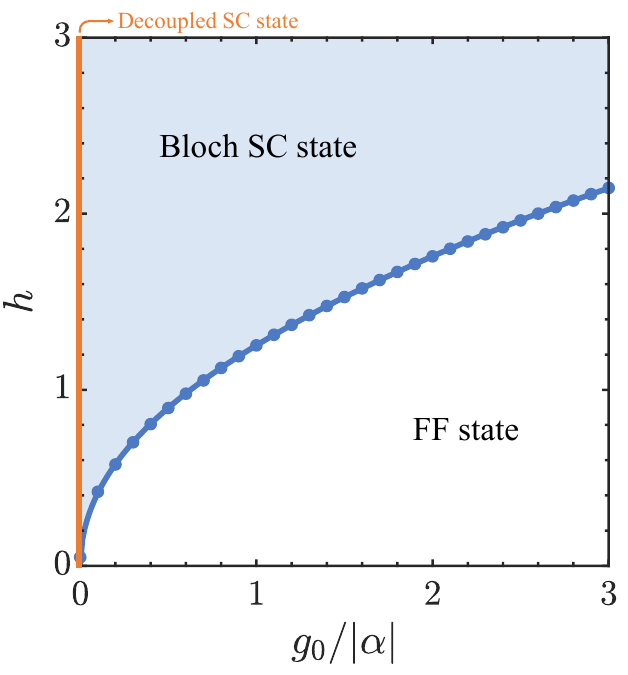}}
			\subfigure[]{
				\includegraphics[width=0.30\linewidth]{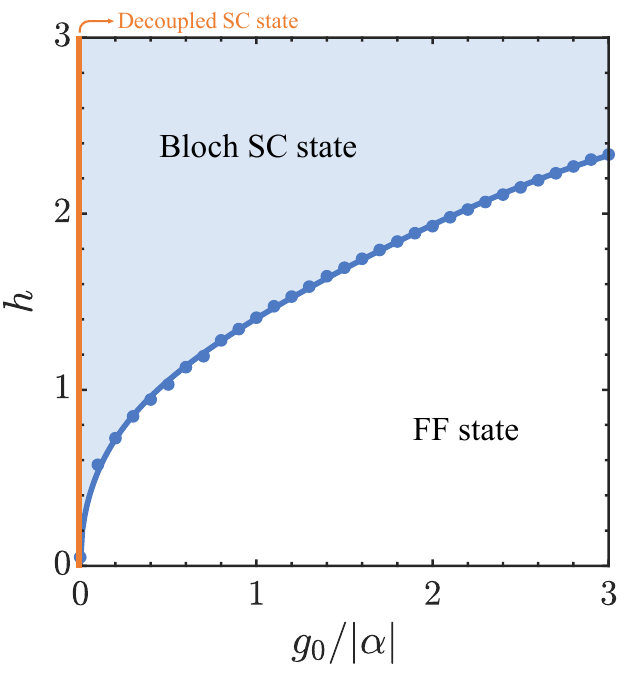}}
			\subfigure[]{
				\includegraphics[width=0.30\linewidth]{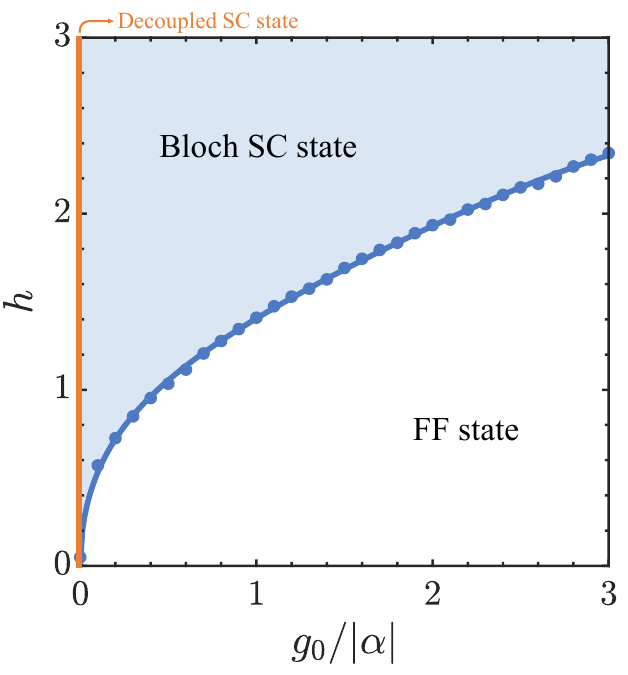}}
			\par\medskip
			\subfigure[]{
				\includegraphics[width=0.30\linewidth]{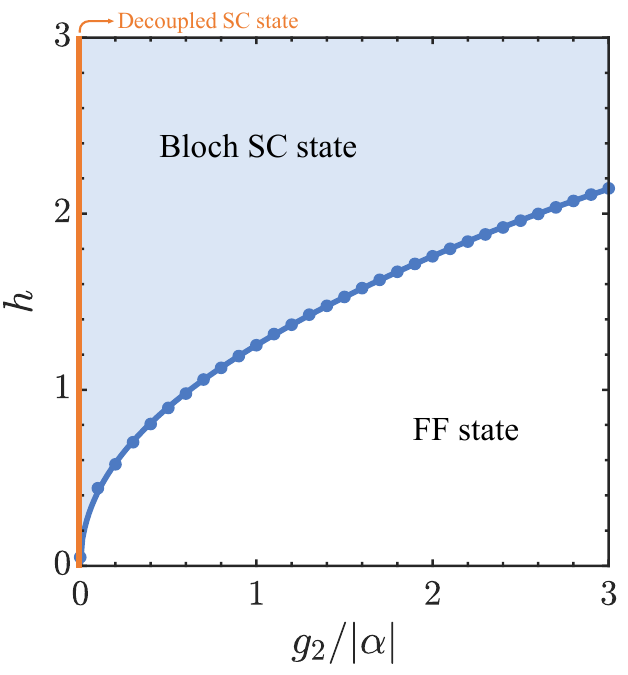}}
			\subfigure[]{
				\includegraphics[width=0.30\linewidth]{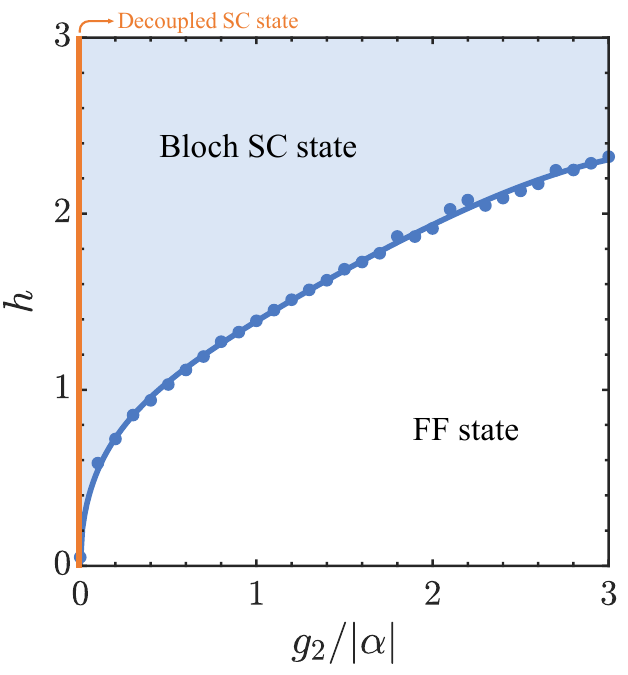}}
			\subfigure[]{
				\includegraphics[width=0.30\linewidth]{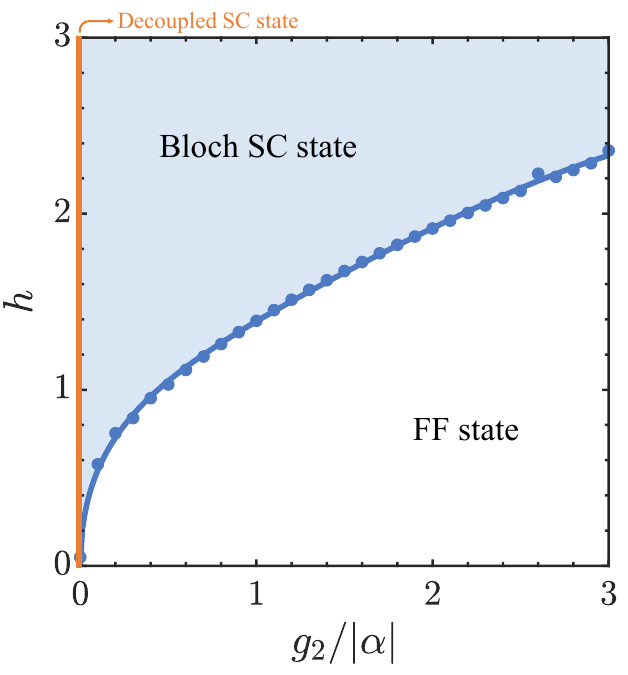}}
			\caption{2D representation: Phase diagram as a function of the Josephson couplings $g_0, g_2$ and the effective field $h$. The values of $\beta$ are: (a,d) $\beta_1=3,\beta_2=1,\beta_3=1$; (b,e) $\beta_1=3,\beta_2=1,\beta_3=-1$; (c,f) $\beta_1=3, \beta_2=-1,\beta_3=1$. The inter-layer coupling $g_2=0$ in the upper panel, while $g_0=0$ in the lower panel.  
				\label{phase2D}}
		\end{center}
	\end{figure}
	
	\section{GL free energy and results for the 1D representations}
	\label{app:1D}
	
	For the four 1D representations $\Gamma_{1,2,3,4}$, the SC order parameter consists of a single component $\psi_l$ per layer. Taking advantage of the periodicity $\psi_l(\mathbf{r}) = \psi_{l+4}(\mathbf{r})$, the free energy density can be reduced to a four-component system ($l = 0, 1, 2, 3$):
	\begin{equation}\label{eq:free}
		f_s - f_n = \sum_{l=0}^{3} \left\{ \frac{\hbar^2}{2m^*} \left| \mathbf{D}_l\psi_l(\mathbf{r}) \right|^2 + \alpha \left| \psi_l(\mathbf{r}) \right|^2 + \frac{\beta}{2} \left| \psi_l(\mathbf{r}) \right|^4 
		- g \left[ \psi_l^*(\mathbf{r}) \psi_{l+1}(\mathbf{r}) + \text{c.c.} \right] \right\}.
	\end{equation}
	The last term represents the Josephson tunneling, with $g$ as the coupling energy. Note that the Peierls phase factor between layers is trivially unity under the gauge choice in Eq.~\eqref{eq:Al}.
	
	The GL equations for the 1D representations are:
	\begin{equation}\label{eq:GLE1}
		-\frac{\hbar^2}{2m^*} \mathbf{D}_l^2\psi_l + \left( \alpha + \beta |\psi_l|^2 \right) \psi_l - g (\psi_{l-1} + \psi_{l+1}) = 0.
	\end{equation}
	
	\subsection{Decoupled and Bloch SC states}
	
	Without Josephson coupling ($g = 0$), the GL equations decouple across layers, yielding a plane-wave solution known as the decoupled SC state~\cite{Qiu2022}:
	\begin{equation}\label{eq:decoupledSC1}
		\psi_l(\mathbf{r}) = \sqrt{\rho_s} e^{i k_0 \hat{n}_l \cdot \mathbf{r}},
	\end{equation}
	up to a layer-dependent phase, where $\rho_s = -\alpha / \beta$ is the uniform superfluid density. Note that the layer-dependent momentum shift $e^{i k_0 \hat{n}_l \cdot \mathbf{r}}$ inherently satisfies the $P_{c}4_{1}22$ symmetry operations.
	
	With finite Josephson coupling ($g \neq 0$), the phase locking between layers competes with the layer-dependent effective gauge field. Consequently, the order parameter tends to modulate within a magnetic unit cell of size $a_0 \times a_0 \times 4a$. We adopt a Bloch wave-function ansatz for this ``Bloch SC state'' (or PDW state):
	\begin{equation}\label{eq:BlochSC1}
		\psi_l(\mathbf{r}) = \varphi_l(\mathbf{r}) e^{i k_0 \hat{n}_l \cdot \mathbf{r}},
	\end{equation}
	where $\varphi_l(x + a_0, y) = \varphi_l(x, y + a_0) = \varphi_l(x, y)$ is a 2D periodic function, expanded as
	\begin{equation}
		\varphi_l(\mathbf{r}) = \sqrt{n} \sum_{\mu, \nu} a_{l,\mu,\nu} e^{i k_0 (\mu x + \nu y)},
	\end{equation}
	with integer indices $\mu, \nu$, and $n$ representing the average superfluid density. The coefficients $a_{l,\mu,\nu}$ are normalized such that $\sum_{\mu,\nu} |a_{l,\mu,\nu}|^2 = 1$.
	
	Minimizing the free energy yields the phase diagram for the 1D representation (Fig.~\ref{figSM_1Dphase}), parameterized by the dimensionless effective-field variable $h$ defined above and the inter-layer coupling strength $|g| / |\alpha|$. Without coupling ($g = 0$), the decoupled state prevails. For finite $g$, a weak effective field favors the FF state, characterized by a uniform amplitude $|\psi_l(\mathbf{r})| = \sqrt{\rho_s^{\text{FF}}}$ and a single, compromise center-of-mass wavevector across all layers. At higher effective fields, the system transitions to the Bloch SC state, consistent with bilayer results~\cite{Qiu2022}.
	
	In the Bloch SC state, the spatial distributions of the order parameter and the associated spontaneous currents are highly intricate. The gauge-invariant intra-layer spontaneous particle current density is given by
	\begin{equation}
		\mathbf{J}_l = \frac{e^* \hbar}{m^*} \text{Im}\left( \psi_l^* \nabla \psi_l \right) - \frac{{e^*}^2}{m^* c} |\psi_l|^2 \mathbf{A}_l,
	\end{equation}
	and the inter-layer Josephson tunneling current is $J_{l,l+1} = -\frac{e^* a}{\hbar} g \operatorname{Im} ( \psi_l^* \psi_{l+1} )$. Note that because $\mathbf{A}_l$ is an effective gauge field rather than a true electromagnetic field, $\mathbf{J}_l$ represents spontaneous local circulating currents---a characteristic vestigial signature of PDW states---rather than Meissner screening currents.
	
\begin{figure*}[h]
		\begin{center}
			\subfigure[]{
				\includegraphics[height=0.28\linewidth]{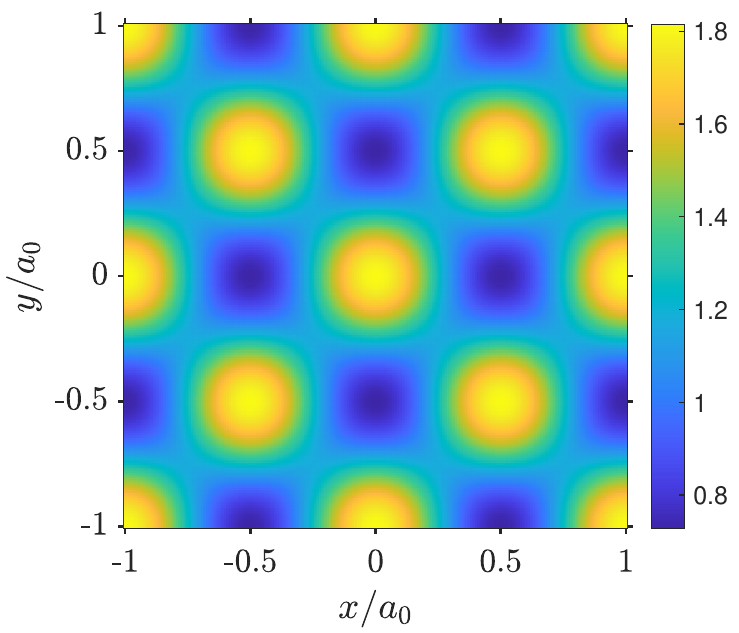}}
			\subfigure[]{
				\includegraphics[height=0.28\linewidth]{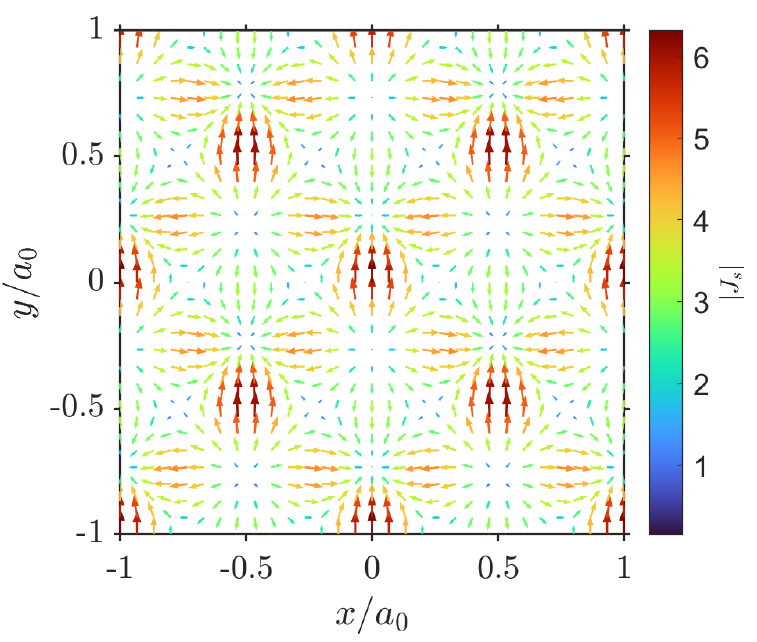}}
			\subfigure[]{
				\includegraphics[height=0.28\linewidth]{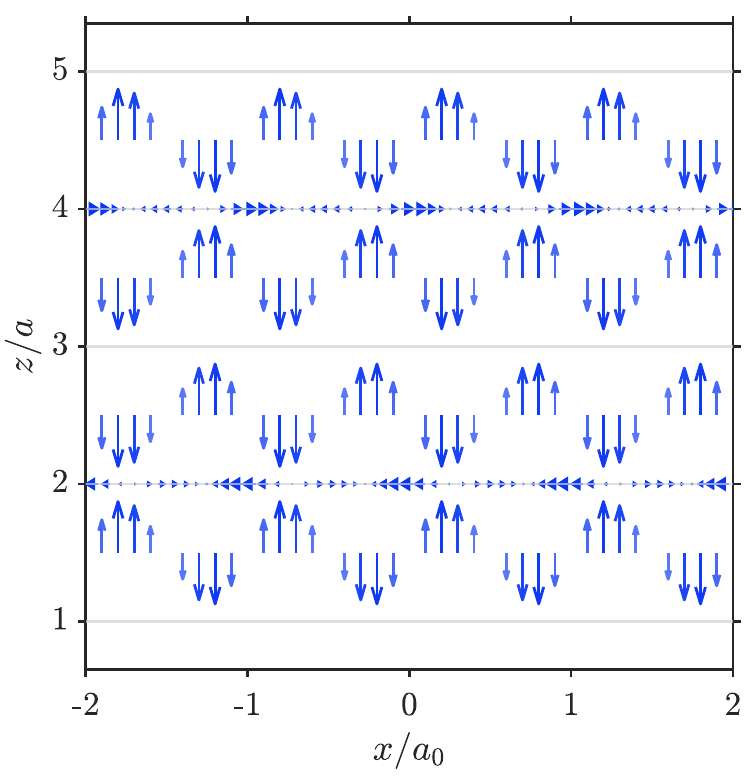}}
			\hfill
			\caption{1D representation: Bloch SC state at $h=2.5$ and $|g|/|\alpha|=1.0$. (a) The superfluid density. (b) The intra-layer spontaneous loop current distribution on the layer $l=0$. (c) The inter-layer Josephson tunneling current is plotted in the xz plane at $y=0$. 
				\label{figSM_1D_Bloch}}
		\end{center}
	\end{figure*}

	Consider a representative Bloch state at $h = 2.5$ and $|g| / |\alpha| = 1.0$. Fig.~\ref{figSM_1D_Bloch}(a) shows the superfluid density $|\psi_l(\mathbf{r})|^2$, exhibiting a checkerboard-like $k_0$-modulated pattern in both $x$ and $y$, indicative of a 2D PDW. The intra-layer spontaneous current and inter-layer Josephson current are depicted in Figs.~\ref{figSM_1D_Bloch}(b) and (c), respectively.
	
	The Fourier-space form used in the minimization follows by inserting Eq.~\eqref{eq:BlochSC1} into Eq.~\eqref{eq:free} and applying the constraints in Eq.~\eqref{eq:aluv}.  With real $l=1$ coefficients $a_{\mu,\nu}\equiv a_{1,\mu,\nu}$ and layer labels understood modulo four, the reduced functional is
	\begin{equation}
		\label{betaF}
		\begin{aligned}
			\beta F =& 4\tilde{n} \sum_{\mu\nu}\left[\frac{\hbar^2 k_0^2}{2 m^*}\left(\mu^2+\nu^2\right)+\alpha\right]a_{\mu,\nu}^2
			+2\tilde{n}^2 \sum_{\substack{\mu_1\mu_2\mu_3 \\ \nu_1\nu_2\nu_3}} a_{\mu_1,\nu_1} a_{\mu_2,\nu_2} a_{\mu_3,\nu_3} a_{\mu_1+\mu_2-\mu_3,\nu_1+\nu_2-\nu_3}
			-8 \tilde{n} g \eta \sum_{\mu\nu} a_{\mu,\nu} a_{\nu-1, -\mu-1}.
		\end{aligned}
	\end{equation}
	where $\tilde{n}=\beta n$
	
	\section{Intra-layer spontaneous loop currents for the 2D representation}
	\label{app:intralayer_currents}
	In the main text, we focused on the superfluid density and the inter-layer Josephson currents. Below, we present the corresponding intra-layer spontaneous loop currents for the three representative Bloch SC states (Diagonal, Principal, and Chiral) discussed in the main text. Note that for the principal nematic and chiral states, the unidirectional current directions rotate by exactly $90^\circ$ between adjacent superconducting layers, providing a direct macroscopic reflection of the microscopic helical magnetic lattice.
	
	\begin{figure*}[h]
		\begin{center}
			\subfigure[Diagonal ($l=0$)]{\includegraphics[width=0.32\linewidth]{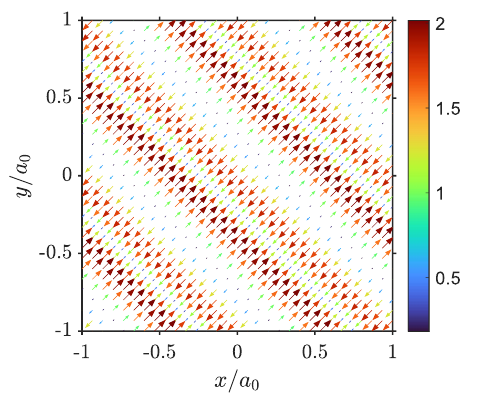}}
			\subfigure[Principal ($l=0$)]{\includegraphics[width=0.32\linewidth]{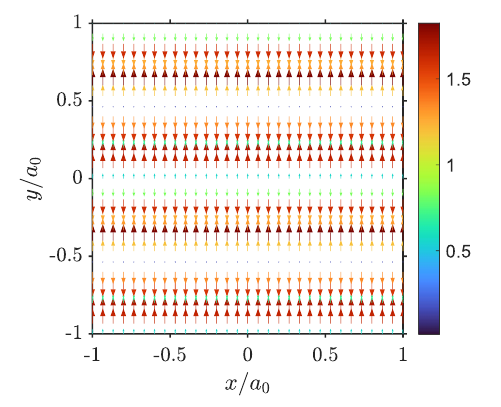}}
			\subfigure[Chiral ($l=0$)]{\includegraphics[width=0.32\linewidth]{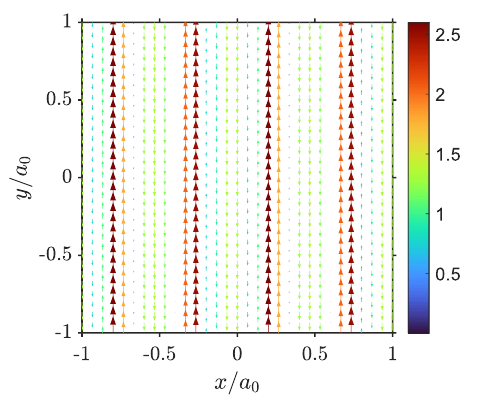}}
			
			\subfigure[Diagonal ($l=1$)]{\includegraphics[width=0.32\linewidth]{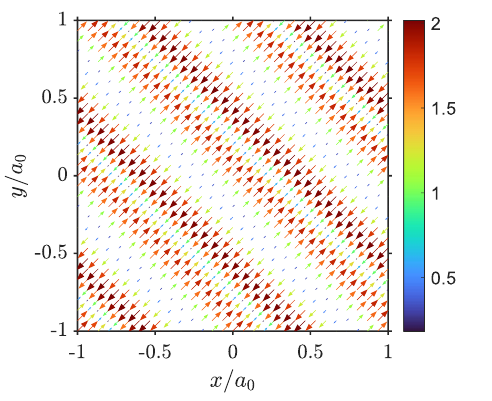}}
			\subfigure[Principal ($l=1$)]{\includegraphics[width=0.32\linewidth]{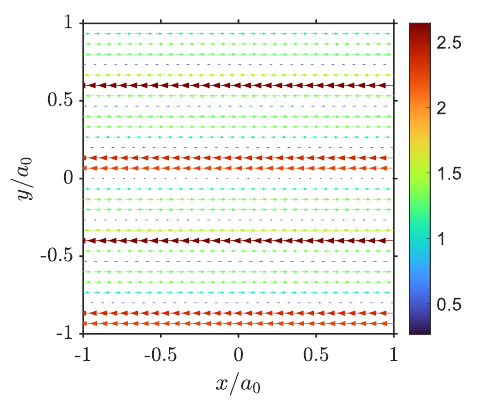}}
			\subfigure[Chiral ($l=1$)]{\includegraphics[width=0.32\linewidth]{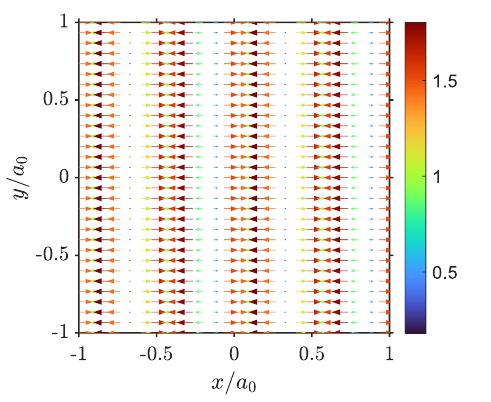}}
			
			\caption{Intra-layer spontaneous loop currents for the corresponding Bloch SC states shown in Fig.~\ref{fig:density} of the main text. The top row displays layer $l=0$ and the bottom row displays the adjacent layer $l=1$. For the principal nematic and chiral states, the unidirectional current directions rotate by exactly $90^\circ$ between adjacent layers, reflecting the underlying helical exchange field. Parameters for each column correspond identically to those in Fig.~\ref{fig:density}(a)--(c).}
			\label{figSM_Intralayer_Currents}
		\end{center}
	\end{figure*}
	
\end{document}